\documentclass[12pt]{article}
\usepackage[dvips]{color}
\usepackage{epsfig}
\usepackage{amsmath}
\usepackage{graphicx}

\begin{document}
\title{\bf  Interacting quintessence dark energy models in Lyra manifold}
\author{{M. Khurshudyan$^{a}$\thanks{Email:
khurshudyan@yandex.ru}, J. Sadeghi$^{b}$\thanks{Email: pouriya@ipm.ir}, R. Myrzakulov$^{c}$\thanks{Email: rmyrzakulov@gmail.com}, A. Pasqua$^{d}$\thanks{Email: toto.pasqua@gmail.com} and H. Farahani$^{e}$\thanks{Email: h.farahani@umz.ac.ir}}\\
$^{a}${\small {\em Department of Theoretical Physics, Yerevan State
University,}}\\
{\small {\em 1 Alex Manookian, 0025, Yerevan, Armenia}}\\
$^{b}${\small {\em Department of Physics, Ayatollah Amoli Branch, Islamic Azad University, Amol, Iran}}\\
$^{c}${\small {\em Eurasian International Center for Theoretical Physics, Eurasian National University,}}\\
{\small {\em Astana 010008, Kazakhstan}}\\
$^{d}${\small {\em Department of Physics, University of Trieste, Via Valerio, 2 34127 Trieste, Italy}}\\
$^{e}${\small {\em Department of Physics, Mazandaran University, Babolsar, Iran}}}  \maketitle
\begin{abstract}
In this paper, we consider two-component dark energy models in Lyra manifold. The first component is assumed as a quintessence field while the second component may be a viscous polytropic gas, a viscous Van der Waals gas or a viscous modified Chaplygin gas. We also consider the possibility of interaction between components. By using the numerical analysis, we study some cosmological parameters of the models and compare them with observational data.\\\\
\noindent {\bf Keywords:} Dark Energy, Cosmology.
\end{abstract}

\section{\large{Introduction}}
Observations of high redshift type Supernovae Ia (SNeIa) \cite{Riess}-\cite{Amanullah} reveal the accelerated expansion of our universe, which nature is not exactly clear until now. It is found that the density of matter is very much less than critical density \cite{Pope}. Moreover, Cosmic Microwave Background (CMB) radiation anisotropies observations indicate that the universe can be considered flat and the total energy density is very close to the critical value ($\Omega_{\small{tot}} \simeq1$) \cite{Spergel}. Based on the experimental data, a component of the energy dubbed as dark energy is thought to be responsible of the physics of the accelerated expansion but it seems that it is not alone in the universe, so the mysterious matter component which is called dark matter should also exist. Dark energy can be described by a pressure sufficiently negative in order to drive the acceleration of the universe and by positive energy density. There are several different models proposed to explain the nature of dark energy. The cosmological constant $\Lambda$ is the simplest model which can be considered, but in presence of many research papers in these fields, the origin of dark energy and dark matter is still unknown, and the possible connection between them is also unknown as well as real role of the components to the history of the universe. This situation gives a lot of freedom to researchers and possibility of some simulations. The cosmological constant faced with two main problems, i.e. the absence of a fundamental mechanism which sets the cosmological constant zero or very small value (which is known as fine-tuning problem) and the problem known as cosmological coincidence problem, which asks why are we living in an epoch in which the densities of dark energy and matter are comparable. One of the interesting way to solve above mentioned problems is to consider interactions between components [6]. From observational point of view, no piece of evidence has been so far presented against such interactions. Indeed, possible interactions between the components of universe have been discussed in recent years. It is found that a suitable interaction can help to alleviate the coincidence problem. Different interacting models of dark energy have been investigated [7-14].\\
Alternative models of dark energy suggest a dynamical form of dark energy, which at least in an effective level, can originate from a variable cosmological constant \cite{Sola}, or from various fields, such is a canonical scalar field \cite{Ratra}-\cite{Saridakis0} (quintessence), a phantom field \cite{Caldwell}-\cite{Dutta}, or quintom \cite{Feng}-\cite{Qiu}. Finally, an interesting attempt to probe the nature of dark energy according to some basic quantum gravitational principles are the holographic dark energy paradigm \cite{Hsu}-\cite{hde} and agegraphic dark energy models \cite{Cai1,WeiCai2}. Among them a quintessence model is interesting in this paper as a component of dark energy. In that case, the dark energy may be dynamical approaching zero with time, or it may be slowly
increasing. It is now dominating the universe because the reduction of mass and radiation
energy density with the scale factor (which gives some information about the size of the universe) is greater than the decrease in
dark energy density in the present epoch. In general, we would like the quintessence field to be decreasing with the scale factor and time at a smaller rate than the mass energy so that it will become dominant at redshifts less than one. The quintessence field has the property of being very weakly coupled to baryons but contributing a negative pressure to the equation of state. In the past it had a small contribution but with time it has decreased less quickly with the scale factor than the matter and radiation densities and is
dominant now.\\
For the, dark energy component we consider several models in this paper, including viscosity. Indeed, bulk viscosity is added to obtain more realistic models. However, a viscous pressure can itself play the role
of an agent that drives the present acceleration of the universe \cite{63}.\\
One of interesting dark energy models is the polytropic gas which was proposed to explain the accelerated
expansion of the universe \cite{64}. It was shown that the polytropic gas model in the
presence of interaction can behave as phantom field \cite{65}. It was pointed out that a
polytropic scalar field can be reconstructed according to the evolutionary behaviors of the holographic and new agegraphic dark energy densities. The validity of the generalized second law
of thermodynamics was also examined for the polytropic gas model in \cite{66}. Another interesting model of dark energy may be Van der Waals gas which could be accounted as a fluid with unusually EoS or could be thought
a fluid satisfying to more general form of EoS i.e. $F(\rho, P) = 0$ \cite{67}. There are also some important models to describe dark energy based on Chaplygin gas equation of state which recently considered by several papers such as \cite{68,69,71,72}, and yield to good agreement with observational data.\\
On the other hand, the Lyra geometry provides one of the possible alternatives in modification of the cosmological models.
As we know the modification of the gravitational theory has long
been famous, but the late-time cosmological acceleration caused to more research in this field \cite{73}.
Now, we like to consider a universe filled with a two-component dark energy in Lyra manifold with possibility of interaction between component. The first component assumed as quintessence, while we have several choices for the second component as viscous polytropic gas, viscous Van der Waals gas or viscous Chaplygin gas.
We suggest these as toy models to describe universe and compare our results with observational data to choose one of them as the best model.\\
This paper is organized as follow. In Section 2,  we introduce our models. In Section 3, we recall the main properties of field equations. In Section 4, we give numerical results corresponding to constant $\Lambda$. In Section 5, we give numerical results corresponding to varying $\Lambda$. In Section 6, we obtain some observational constraints. Finally, in Section 7, we write the Conclusions of this paper.
\section{\large{The models}}
One of the well studied Dark Energy models is the quintessence model \cite{Ratra}-\cite{Kh}, which is a scalar field model described by a field $\phi$ and a $V(\phi)$ potential. It represents the simplest scalar-field scenario without having theoretical problems such as the appearance of ghosts and Laplacian instabilities. The energy density $\rho_Q$ and the  pressure $P_Q$ of the quintessence scalar field model are given, respectively, by:
\begin{equation}\label{eq:rhoQ}
\rho_{Q}=\frac{1}{2}\dot{\phi}^{2}+V(\phi),
\end{equation}
and,
\begin{equation}\label{eq:rhoP}
P_{Q}=\frac{1}{2}\dot{\phi}^{2}-V(\phi).
\end{equation}
Canonical scalar field is not the unique solution. We can generalize it as follow \cite{74},
\begin{equation}\label{eq:rhoQnew}
\rho_{Q}=\frac{\omega}{2}\phi^{k}\dot{\phi}^{2}+V(\phi),
\end{equation}
and,
\begin{equation}\label{eq:rhoPnew}
P_{Q}=\frac{\omega}{2}\phi^{k}\dot{\phi}^{2}-V(\phi).
\end{equation}
In the case of $k = 0$, Eqs. (\ref{eq:rhoQnew}) and (\ref{eq:rhoPnew}) transform to the canonical scalar field model with a rescaling of the field.
Below, we would like to consider an interaction term $Q$ between dark energy and dark matter described by,
\begin{equation}\label{eq:Q}
Q=3Hb\rho_{Q}+\gamma (\rho_{b}-\rho_{Q})\frac{\dot{\phi}}{\phi},
\end{equation}
where $b$ and $\gamma$ are positive constants with a typical value of $0.01-0.03$.  Nature of the interaction between dark energy and dark matter is not clear. If we believe that it has a quantum origin, then an absence of the final theory of quantum gravity leave this question as an open problem. However if we believe that the link exist between components is due to the same origin of the dark energy and dark matter, then this approach does not give any exact solution, because the nature of the two component is not formulated and it is an other open problem. Therefore, only phenomenological assumption is an appropriate approach.
For the dark matter model we will consider once a viscous modified Chaplygin gas with the following equation of state (EoS),
\begin{equation}\label{eq:vChgas}
P=A\rho-\frac{B}{\rho^{\alpha}}-3\xi H,
\end{equation}
where $A$, $B$ and $\alpha$ are constants (with $0 \leq \alpha \leq 1$ in General Relativity).\\
For the second model we will use viscous polytropic fluid with EoS given by:
\begin{equation}\label{eq:PolEoS}
P=K\rho^{1+1/n}-3\xi H,
\end{equation}
where $K$ is the polytropic index and $\xi$ represents the viscous coefficient.\\
In the third model we would like to consider interaction between quintessence dark energy and a viscous van der Waals gas of the general form:
\begin{equation}\label{eq:WaalsGas}
P=\frac{A\rho}{B-\rho}-B\rho^{2}-3\xi H,
\end{equation}
where $A$ and $B$ are constants. Furthermore, we will consider two regimes: 1) when $\Lambda$ is a numerical constant; 2) $\Lambda$ is a function of the cosmic time $t$, therefore it is a varying quantity. In particular, we choose the following form for the time-varying $\Lambda$:
\begin{equation}\label{eq:lambda}
\Lambda(t)=H^{2}\phi^{-2}+\delta V(\phi),
\end{equation}
where $\delta$ is a positive constant, $V(\phi)$ is the potential of the field which we consider as follow,
\begin{equation}\label{eq:pot}
V(\phi)=V_{0}e^{\left(-\phi_{0}\phi\right)},
\end{equation}
where $\phi_0$ is a constant parameter.\\
We investigate the behavior of the cosmological parameters like the Hubble parameter $H$, deceleration parameter $q$, EoS parameters of the quintessence dark energy and an effective two component fluid. Moreover, we perform stability analysis via the squared speed of the sound $C_S^2$, which is defined as follow:
\begin{equation}\label{eq:Cstab}
C_{s}^{2}=\frac{\partial{P}}{\partial{\rho}},
\end{equation}
where $P$ and $\rho$ are, respectively, the pressure and the energy density of the effective fluid given by:
\begin{equation}
P=P_{Q}+P_{i},
\end{equation}
and,
\begin{equation}
\rho=\rho_{Q}+\rho_{i},
\end{equation}
where $i$ refers to one of the viscous fluids described above. We will finish our paper with the results obtained from observational constraints. Consideration of the statefinder analysis, different forms of interaction terms as well as different $\Lambda(t)$ and varying viscosity is possible and an interesting research. We hope to approach to these question in future with forthcoming articles.
\section{\large{The field equations}}
The field equations governing our model are given by [73],
\begin{equation}\label{eq:Einstein eq}
R_{\mu\nu}-\frac{1}{2}g_{\mu\nu}R-\Lambda g_{\mu \nu}+\frac{3}{2}\phi_{\mu}\phi_{\nu}-\frac{3}{4}g_{\mu \nu}\phi^{\alpha}\phi_{\alpha}=T_{\mu\nu}.
\end{equation}
Considering the content of the Universe to be a perfect fluid, we have,
\begin{equation}\label{eq:T}
T_{\mu\nu}=(\rho+P)u_{\mu}u_{\nu}-Pg_{\mu \nu},
\end{equation}
where $u_{\mu}=(1,0,0,0)$ is the 4-velocity of the co-moving
observer, satisfying  the relation $u_{\mu}u^{\mu}=1$. Let $\phi_{\mu}$ be a time-like
vector field of displacement, then:
\begin{equation}
\phi_{\mu}=\left ( \frac{2}{\sqrt{3}}\beta,0,0,0 \right ),
\end{equation}
where $\beta=\beta(t)$ is a function of time alone, and the factor $2/\sqrt{3}$ is inserted in order to simplify the writing of all the following equations.
By using FRW metric for a flat Universe given by:
\begin{equation}\label{s2}
ds^2=-dt^2+a(t)^2\left(dr^{2}+r^{2}d\Omega^{2}\right),
\end{equation}
field equations can be reduced to the following Friedmann equations,
\begin{equation}\label{eq:f1}
3H^{2}-\beta^{2}=\rho+\Lambda,
\end{equation}
and
\begin{equation}\label{eq:Freidmann2}
2\dot{H}+3H^{2}+\beta^{2}=-P+\Lambda,
\end{equation}
where $H=\frac{\dot{a}}{a}$ is the Hubble parameter, and dot
stands for differentiation with respect to the cosmic
time $t$, $d\Omega^{2}=d\theta^{2}+\sin^{2}\theta d\phi^{2}$, and $a(t)$
represents the scale factor. The $\theta$ and $\phi$ parameters are
the usual azimuthal and polar angles of spherical coordinates, with
$0\leq\theta\leq\pi$ and $0\leq\phi<2\pi$. The coordinates ($t, r,
\theta, \phi$) are called co-moving coordinates.\\
The continuity equation is given by:
\begin{equation}\label{eq:coneq}
\dot{\rho}+\dot{\Lambda}+2\beta\dot{\beta}+3H(\rho+P+2\beta^{2})=0.
\end{equation}
The continuity equation given in Eq. (\ref{eq:coneq}) can be also rewritten in the compact form:
\begin{equation}\label{eq:DEDM}
\dot{\rho}+3H(\rho+P)=0.
\end{equation}
Comparing Eqs. (\ref{eq:coneq})and (\ref{eq:DEDM}) we obtain the following link between $\Lambda$ and $\beta$:
\begin{equation}\label{eq:lbeta}
\dot{\Lambda}+2\beta\dot{\beta}+6H\beta^{2}=0.
\end{equation}
In order to introduce an interaction between dark energy and dark matter, we should mathematically split Eq. (\ref{eq:DEDM}) into the two following equations:
\begin{equation}\label{eq:inteqm}
\dot{\rho}_{i}+3H(\rho_{i}+P_{i})=Q,
\end{equation}
and,
\begin{equation}\label{eq:inteqG}
\dot{\rho}_{Q}+3H(\rho_{Q}+P_{Q})=-Q.
\end{equation}
The cosmological parameters of our interest are the EoS parameters of each fluid components $\omega_{i}=P_{i}/\rho_{i}$, the EoS parameter of composed fluid,
$$\omega_{tot}=\frac{P_{Q}+P_{i} }{\rho_{Q}+\rho_{i}},$$ and the deceleration parameter $q$, which can be written as follow:
\begin{equation}\label{eq:accchange}
q=\frac{1}{2}(1+3\frac{P}{\rho} ),
\end{equation}
where index $i$ refers to the first components may be viscous modified Chaplygin gas or viscous polytropic fluid, and index $Q$ refers to the quintessence scalar field. A differential equation describing dynamics of the DE after some mathematics can be rewritten as,
\begin{equation}\label{eq:rhoQ1}
\dot{\rho}_{Q}+3H\rho_{Q} \left ( 1+b+\omega_{Q} -\frac{\gamma}{3H} \frac{\dot{\phi}}{\phi} \right )=-\gamma \rho_{i}\frac{\dot{\phi}}{\phi}.
\end{equation}
Taking into account the form of the varying $\Lambda(t)$ from Eq. (\ref{eq:lambda}) for the Hubble parameter $H$ we will have:
\begin{equation}\label{eq:Hubble}
H=\frac{1}{\sqrt{3}}\sqrt{\frac{\rho+\delta e^{[-\phi_{0}\phi]}+\beta^{2}}{1-\frac{\phi^{-2}}{3}}}.
\end{equation}
Hereafter, we will consider $\phi_{0}=1$ for mathematical simplicity.
\section{\large{Case of constant $\Lambda$}}
We found reasonable to start our analysis from the models with constant $\Lambda$. Without loss of generality, we would like to describe equations allowing us to find dynamics of the models. According to the assumption with constant $\Lambda$ Eq. (20) will be modified as follow:
\begin{equation}
\dot{\rho}+2\beta\dot{\beta}+3H(\rho+P+2\beta^{2})=0.
\end{equation}
and, using the expression $\dot{\rho}+3H(\rho+P)=0$, we will obtain that:
\begin{equation}
\dot{\beta}+3H\beta=0.
\end{equation}
The last equation can be integrated very easily and the result is the following,
\begin{equation}
\beta=\beta_{0}a^{-3},
\end{equation}
where $a(t)$ is the scale factor and $\beta_{0}$ is the integration constant. In our future calculations we will use $\beta_{0}=1$ as initial condition. For the Hubble parameter $H$ we will obtain,
\begin{equation}\label{eq:ClambdaHubble}
H=\frac{1}{\sqrt{3}}\sqrt{\rho+\Lambda+\beta_{0}a^{-6}}.
\end{equation}
Concerning to the form of the field equations, we need only to assume the form of $Q$ and we will obtain the cosmological solutions. Concerning to the mathematical hardness of the problem we will analyze models numerically and investigate graphical behavior of various important cosmological parameters respect to the cosmic time. In the following subsections we consider three models with the particular form of $Q$ with given forms of the EoS for the viscous dark matter fluids considered in introduction section.

\subsection{\large{Model 1}}
The first toy model describe the dynamics of the Universe within an effective fluid in case of the cosmological constant. The dynamics of the energy density of the viscous modified Chaplygin gas which will model dark matter in our Universe and the differential equation describing the dynamics of it can be found to be,
\begin{equation}\label{eq:M1}
\dot{\rho}_{Ch}+3H \left (1+A-\frac{B}{\rho_{Ch}^{\alpha+1}}-\frac{\gamma}{3H}\frac{\dot{\phi}}{\phi} \right )\rho_{Ch}=3H \left ( b-\frac{\gamma}{3H}\frac{\dot{\phi}}{\phi} \right )\rho_{Q}+9\xi H.
\end{equation}
From Eq. (26) for the dynamics of the dark energy, we have:
\begin{equation}
\dot{\rho}_{Q}+3H \left ( 1+b +\omega_{Q} -\frac{\gamma}{3H} \frac{\dot{\phi}}{\phi}\right  ) \rho_{Q}=-\gamma\frac{\dot{\phi}}{\phi}\rho_{Ch}.
\end{equation}
The best fit for the theoretical model of our consideration with the observational data is obtained for $H_{0}=1.4$, $\Omega_{Ch}=0.3$, $\Lambda=0.7$, $A=2.5$, $B=0.9$, $\gamma=0.02$, $b=0.01$ and $\xi=0.1$.\\
In Figs. 1-3, we have chosen to plot some of the quantities derived for different values of the parameters involved. In particular, we have chosen $\gamma= 0.02$, $b=0.01$, $\xi=0.1$, $\alpha = 0.5$, $A=2.5$ and we have chosen five different values of $\Lambda$, i.e. 0, 0.2, 0.3, 0.5, 0.7.\\
Fig. 1 shows that the Hubble parameter $H$ is decreasing with time to a constant at the late universe as expected and its value increased by $\Lambda$. On the other hand, the value of the deceleration parameter decreased with $\Lambda$. It is illustrated by the right plot of the Fig.1. For $\Lambda=0$ we can see $q\rightarrow-0.4$ while $\Lambda=0.7$ we can see $q\rightarrow-0.7$. Also acceleration to deceleration phase transition seen in this model. At the late time we have $q\sim-0.5$ in agreement with observational data. As we know recent observations of type SNIa represent
that universe is accelerating with the deceleration parameter lying somewhere in the range $-1 < q \leq 0$.\\
Moreover, Fig. 2 show that the EoS parameter tends to -1 at the late time with $\omega_{tot}\geq-1$, corresponding to a quintessence-like universe.\\
Unfortunately, analysis of squared sound speed (see Fig. 3) shows that this model is not stable at the late time and will be considered only for the early universe.

\begin{figure}[h!]
 \begin{center}$
 \begin{array}{cccc}
\includegraphics[width=50 mm]{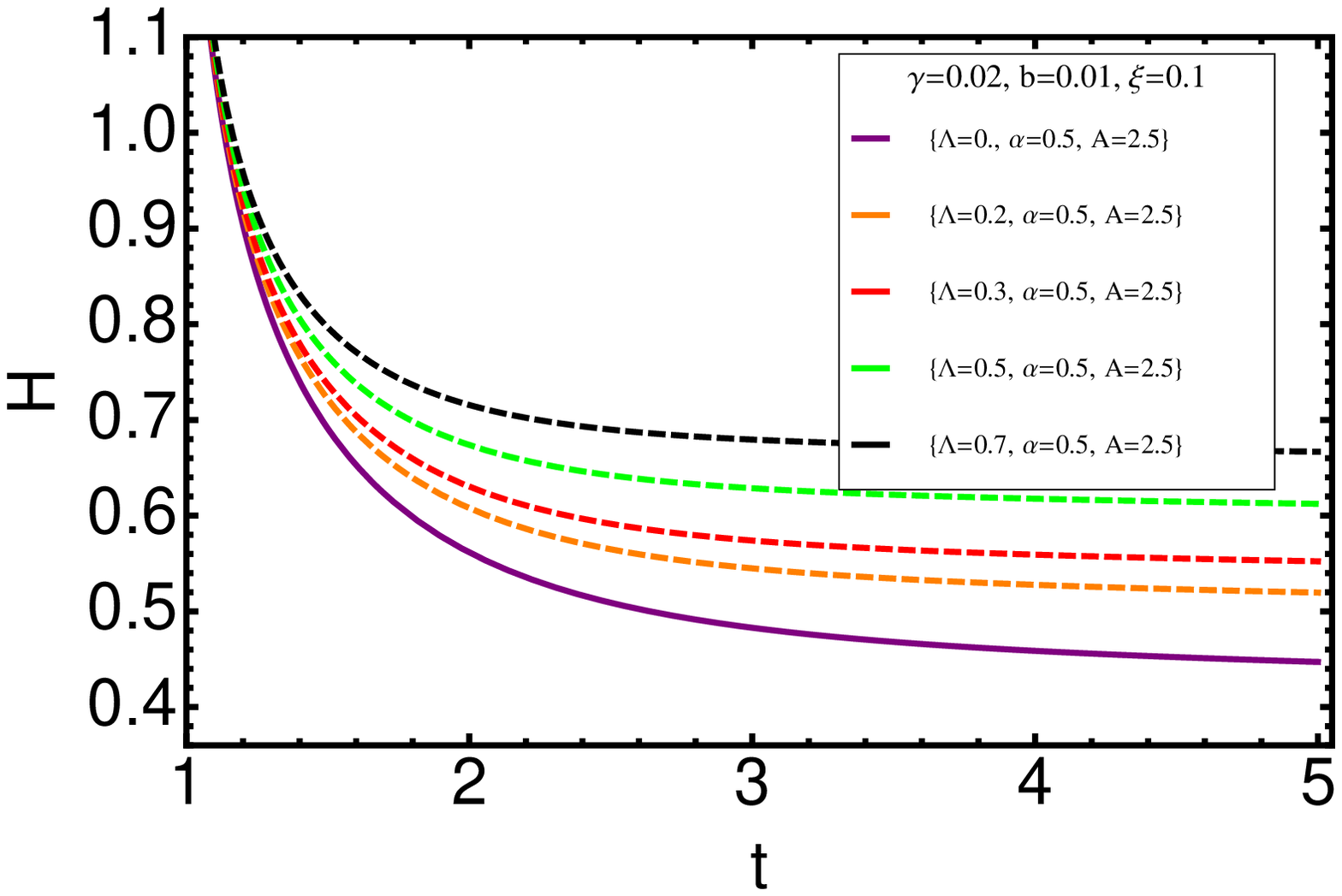} &
\includegraphics[width=50 mm]{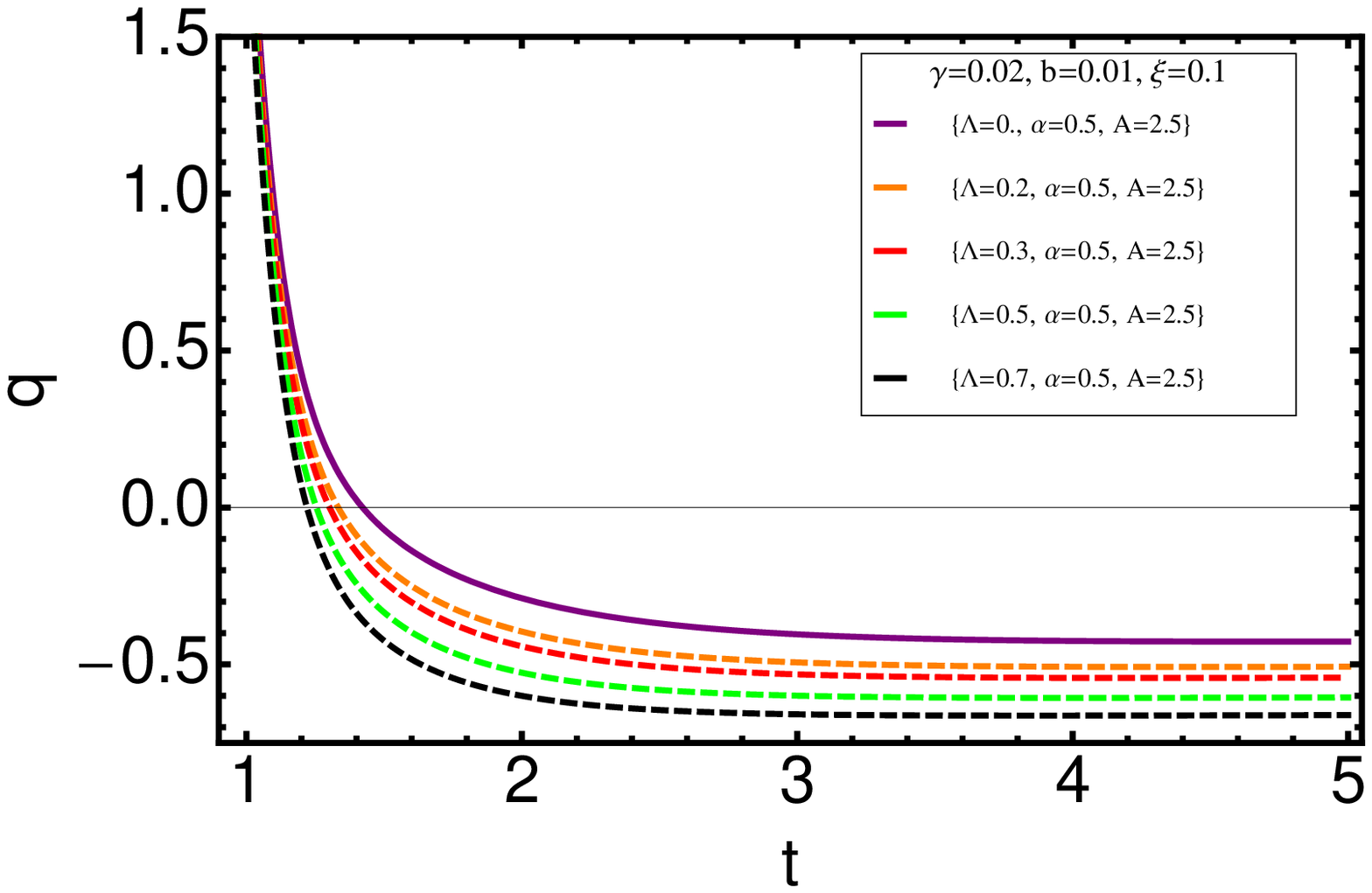}
 \end{array}$
 \end{center}
\caption{Behavior of Hubble parameter $H$ and  $q$ against $t$ for the constant $\Lambda$. Model 1}
 \label{fig:const1}
\end{figure}

\begin{figure}[h!]
 \begin{center}$
 \begin{array}{cccc}
\includegraphics[width=50 mm]{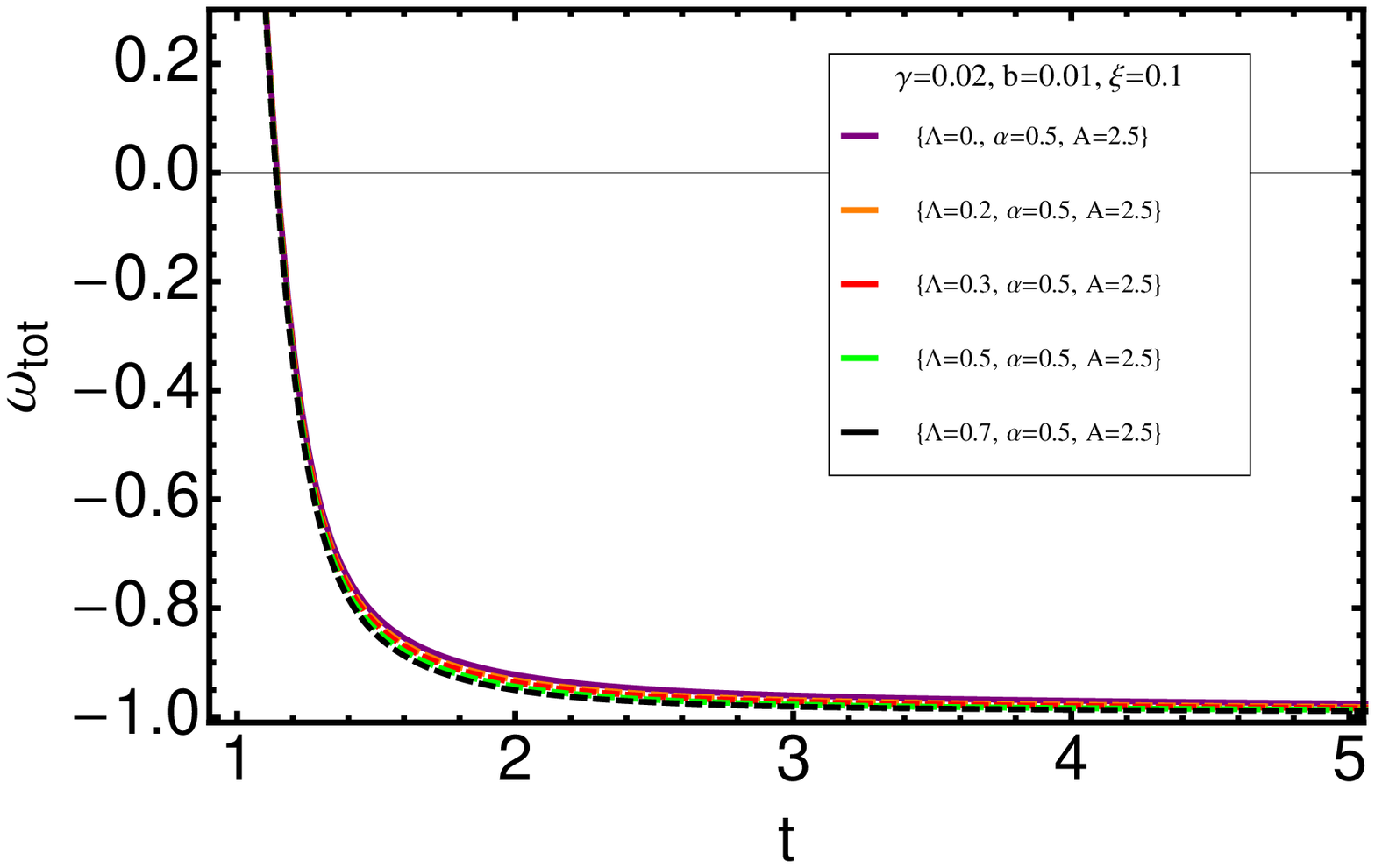} &
\includegraphics[width=50 mm]{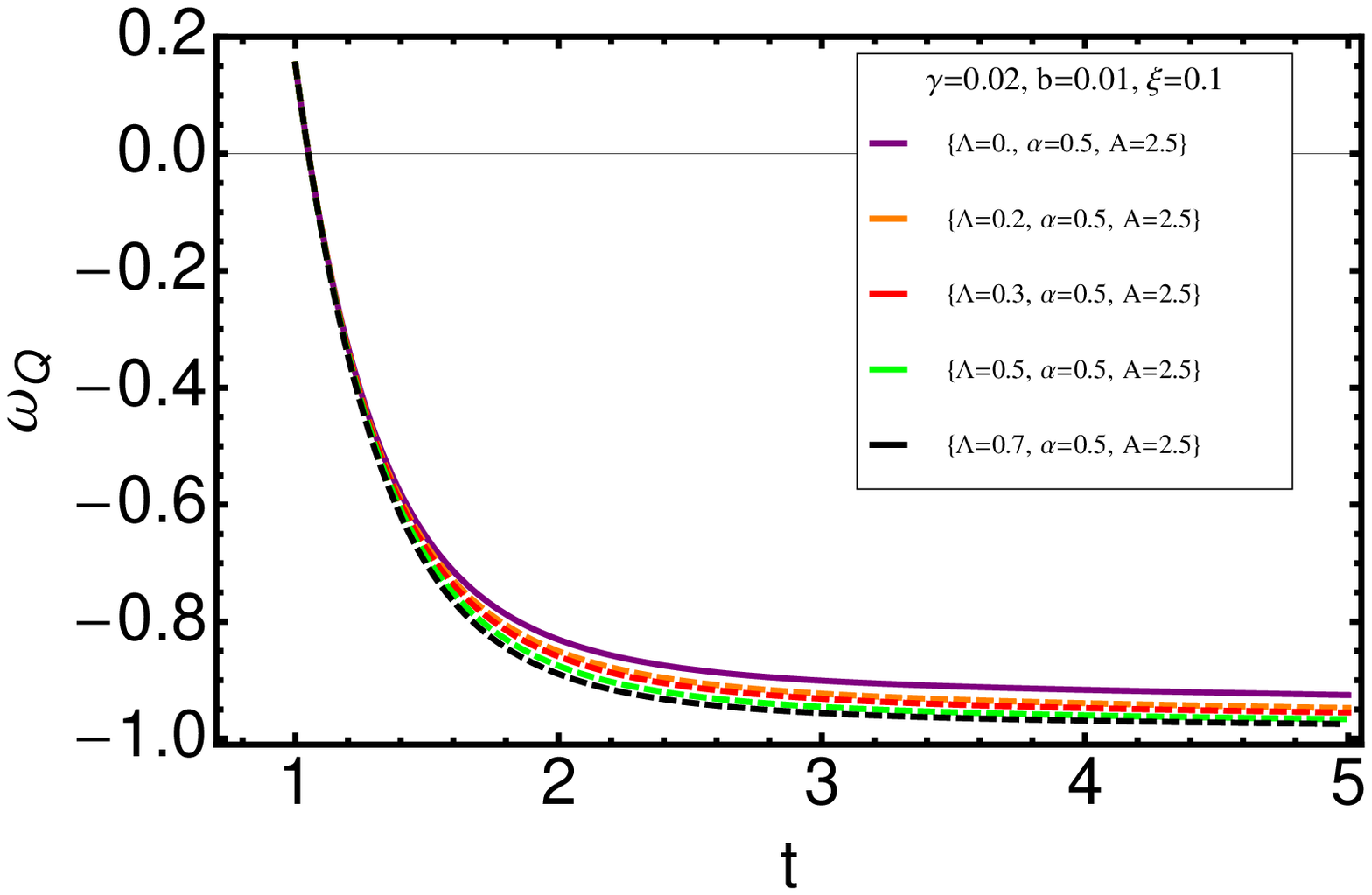}\\
 \end{array}$
 \end{center}
\caption{Behavior of EoS parameter $\omega_{tot}$ and $\omega_{Q}$ against $t$ for the constant $\Lambda$. Model 1}
 \label{fig:const2}
\end{figure}

\begin{figure}[h!]
 \begin{center}$
 \begin{array}{cccc}
\includegraphics[width=50 mm]{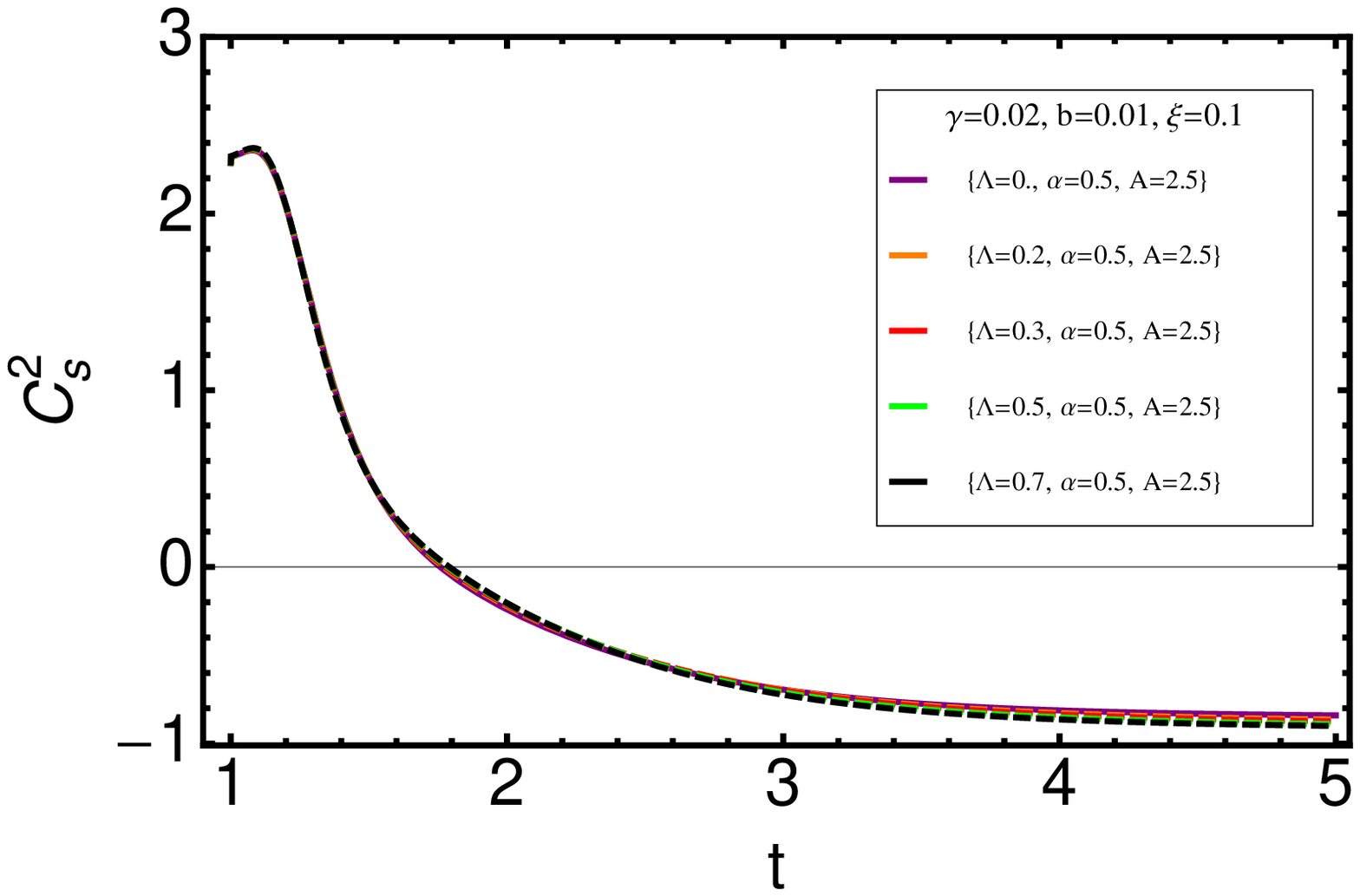}\\
 \end{array}$
 \end{center}
\caption{Squared sound speed against $t$ for the constant $\Lambda$. Model 1}
 \label{fig:const3}
\end{figure}

\subsection{\large{Model 2}}
In the second model, after some mathematical calculations, we obtain the following differential equation to study dynamics of the model,
\begin{equation}\label{eq:M2}
\dot{\rho}_{P}+3H \left (1+\omega_{P}-\frac{\gamma}{3H}\frac{\dot{\phi}}{\phi} \right )\rho_{P}=3H \left ( b-\frac{\gamma}{3H}\frac{\dot{\phi}}{\phi} \right )\rho_{Q}+9\xi H,
\end{equation}
where $\omega_{P}$ given as,
\begin{equation}\label{eq:omegaP}
\omega_{P}=K\rho_{P}^{1/n}-\frac{3\xi H}{\rho_{P}}.
\end{equation}
The best fit for the theoretical model of our consideration with the observational data we obtained for $H_{0}=1.2$, $\Omega_{P}=0.25$, $\Lambda=1.5$, $K=2.5$, $n=1.0$ and $\gamma=0.02$, $b=0.01$, $\xi=0.2$.\\
In Figs. 4 and 5, we have chosen to plot some of the quantities derived for different values of the parameters involved. In particular, we have chosen $\gamma= 0.02$, $b=0.01$, $\xi=0.1$, $n=1$, $K=2.5$ and we have chosen five different values of $\Lambda$, i.e. 0, 0.5, 1.0, 1.2, 1.5.\\
Numerical results of the Hubble expansion parameter and the deceleration parameter $q$ show good behavior (see Fig. 4), but the stability analysis illustrated in Fig. 5 shows that this model also has instability in the late time and is only useful for the early universe. However, the effect of constant $\Lambda$ in these parameter are similar the first model and the EoS parameter tend -1 as before. In the case of $\Lambda=0$, we can see that $q\sim-0.2$, which is not coincident with observational data. It tells that presence of $\Lambda$ may be necessary to obtain agreement with observations.\\
Fig. 5 show that instability of model may solve for the large value of the cosmological constant. Therefore the model 2 is stable at the late time for the large value of the $\Lambda$.

\begin{figure}[h!]
 \begin{center}$
 \begin{array}{cccc}
\includegraphics[width=50 mm]{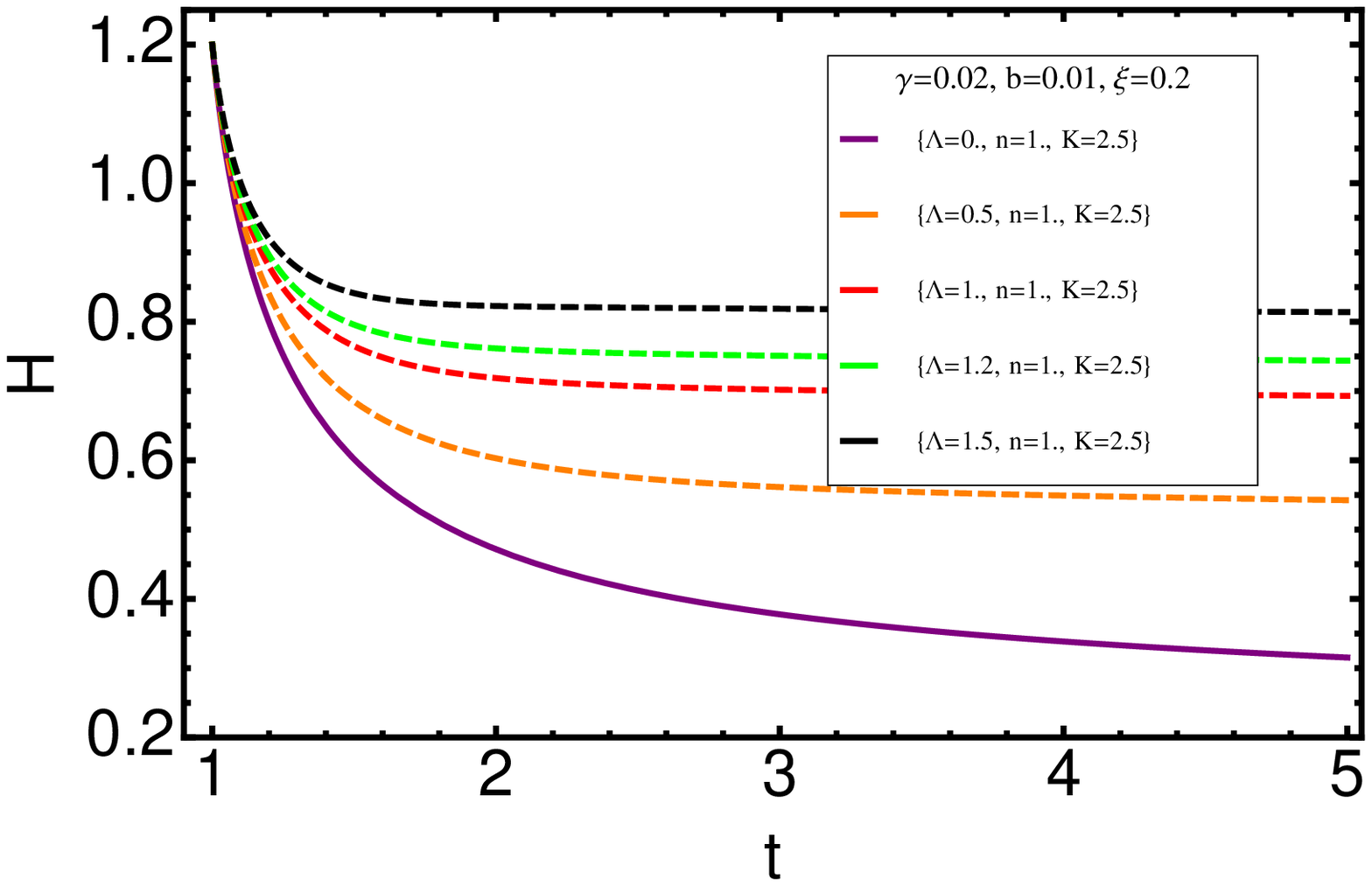} &
\includegraphics[width=50 mm]{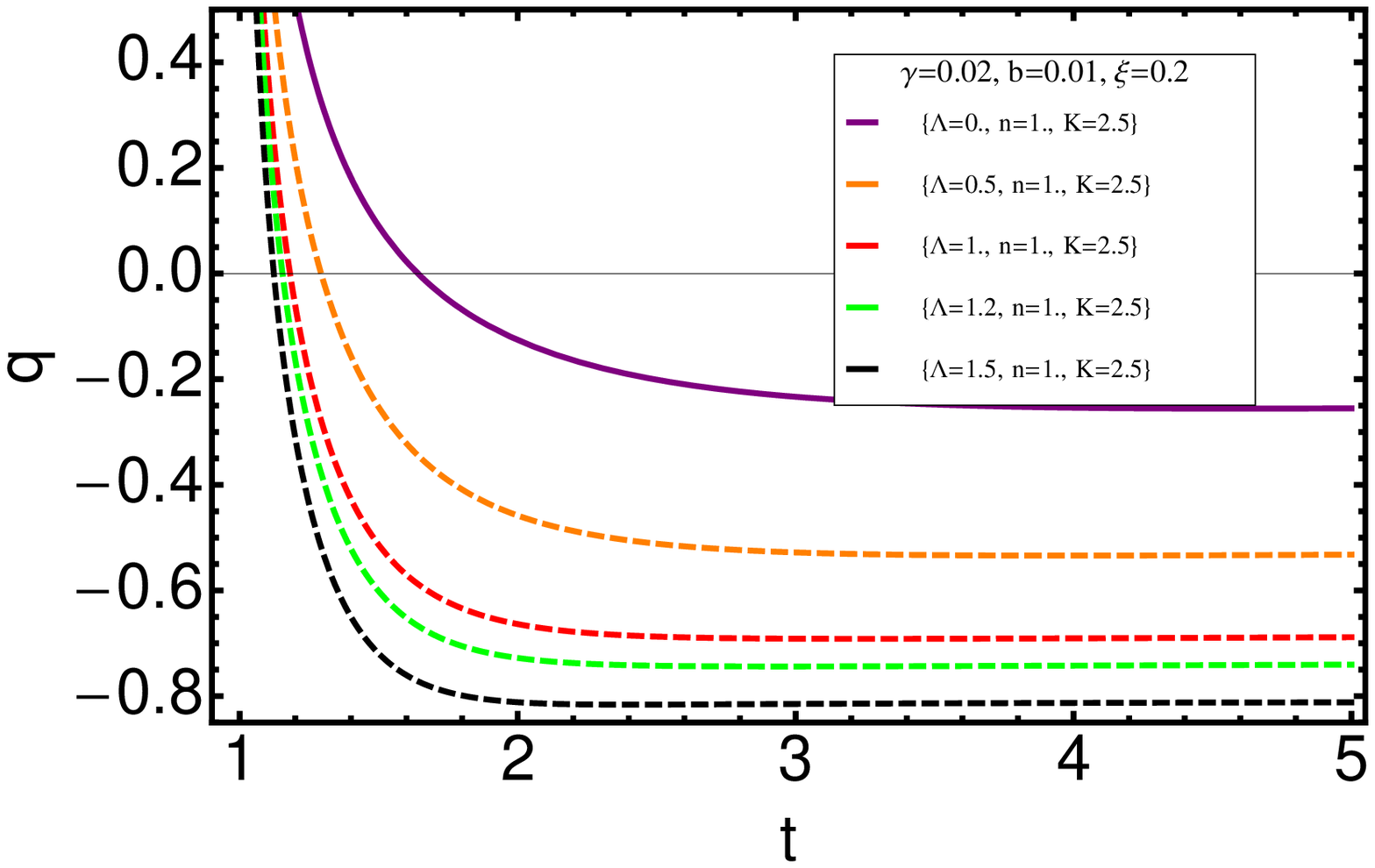}
 \end{array}$
 \end{center}
\caption{Behavior of Hubble parameter $H$ and  $q$ against $t$ for the constant $\Lambda$. Model 2}
 \label{fig:const4}
\end{figure}

\begin{figure}[h!]
 \begin{center}$
 \begin{array}{cccc}
\includegraphics[width=50 mm]{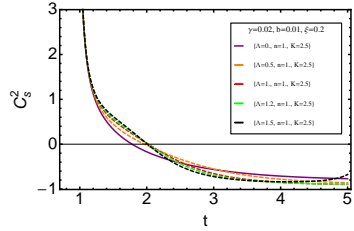}\\
 \end{array}$
 \end{center}
\caption{Squared sound speed against $t$ for the constant $\Lambda$. Model 2}
 \label{fig:const6}
\end{figure}

\subsection{\large{Model 3}}
In the third model we have the following expression for the pressure $P$:
\begin{equation}\label{eq:EoS_M3}
P=\frac{A\rho_{W}}{B-\rho_{W}}-B\rho_{W}^{2}-3\xi H.
\end{equation}
The constant $\Lambda$ assumption will lead us to the following expression for the Hubble parameter $H$,
\begin{equation}\label{eq:H3}
H=\frac{1}{\sqrt{3}}\sqrt{\rho+\Lambda+\beta_{0}a^{-6}}.
\end{equation}
The differential equations describing the dynamics of the energy densities of both components are given by the following equations,
\begin{equation}\label{eq:rhoW}
\dot{\rho}_{W}+3H\rho_{W} \left ( 1+\frac{A}{B-\rho_{W}} -B\rho_{W} -\frac{\gamma}{3H}\frac{\dot{\phi}}{\phi} \right )=3H\rho_{Q}\left ( b-\frac{\gamma}{3H}\frac{\dot{\phi}}{\phi} \right )+9\xi H,
\end{equation}
and,
\begin{equation}\label{eq:rhoQ2}
\dot{\rho}_{Q}+3H\rho_{Q} \left ( 1+b+\omega_{Q} -\frac{\gamma}{3H} \frac{\dot{\phi}}{\phi} \right )=-\gamma \rho_{W}\frac{\dot{\phi}}{\phi},
\end{equation}
where $\omega_{Q}$ is the EoS parameter of dark energy. The best fit for the theoretical model of our consideration with the observational data is obtained for $H_{0}=1.3$, $\Omega_{P}=0.217$, $\Lambda=1.2$, $A=1.5$, $B=1.2$ and $\gamma=0.02$, $b=0.00$, $\xi=0.4$.\\
The behavior of the cosmological parameters are similar to the previous models and we can see late time instability of this model in Fig. 6. This suggest to consider varying $\Lambda$ to obtain more appropriate models.

\begin{figure}[h!]
 \begin{center}$
 \begin{array}{cccc}
\includegraphics[width=50 mm]{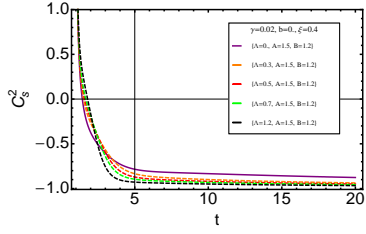}\\
 \end{array}$
 \end{center}
\caption{Squared sound speed against $t$ for the constant $\Lambda$. Model 3}
 \label{fig:const9}
\end{figure}

\section{\large{The case of varying $\Lambda$}}
In this section we will consider three interacting fluid models and will investigate cosmological parameters like the Hubble parameter $H$, deceleration parameter $q$, EoS parameters of the total fluid and dark energy $\omega_{Q}$. Based on numerical solutions, we will discuss graphical behaviors of the cosmological parameters. For the varying $\Lambda$ we take a phenomenological form which was considered by us recently \cite{75}. The formula of $\Lambda$ is given as the following expression,
\begin{equation}
\Lambda(t) = H^{2} \phi^{-2}+\delta V(\phi),
\end{equation}
which is a function of the Hubble parameter, potential of the scalar field, and time derivative of the scalar field. For the potential we take a simple form $V(\phi)=e^{[-\phi]}$, therefore the form of $\Lambda$ can be written also in the following way as only a function of the filed $\phi$,
\begin{equation}
\Lambda(t)=H^{2}\phi^{-2}+\delta e^{-\phi}.
\end{equation}
Therefore, the dynamics of $\beta$ can be obtained from the following differential equation,
\begin{equation}
2\beta \dot{\beta}+6H\beta^{2}+2H\dot{H}\phi^{-2}-2H^{2}\phi^{-3}\dot{\phi}-\delta  e^{[-\phi]} \dot{\phi}=0.
\end{equation}
In forthcoming subsections within three different forms of $Q$ we will investigate the dynamics of the Universe. The question of the dynamics for the energy densities of the dark energy and dark matter is already discussed in previous section, therefore we will not consider them here and we will start with the comments on the graphical behaviors of the cosmological parameters of the models. We will start with the model where,
\begin{equation}
Q=3Hb\rho_{Q}+\gamma (\rho_{i}-\rho_{Q})\frac{\dot{\phi}}{\phi}.
\end{equation}

\subsection{\large{Model 4}}
Interacting viscous modified Chaplygin gas with the quintessence dark energy in the case of varying $\Lambda$ gives the following differential equation,
\begin{equation}\label{eq:M1-1}
\dot{\rho}_{Ch}+3H \left (1+A-\frac{B}{\rho_{Ch}^{\alpha+1}}-\frac{\gamma}{3H}\frac{\dot{\phi}}{\phi} \right )\rho_{Ch}=3H \left ( b-\frac{\gamma}{3H}\frac{\dot{\phi}}{\phi} \right )\rho_{Q}+9\xi H,
\end{equation}
with the Hubble parameter which obtained as,
\begin{equation}\label{eq:Hubble4}
H=\frac{1}{\sqrt{3}}\sqrt{\frac{\rho+\delta e^{[-\phi_{0}\phi]}+\beta^{2}}{1-\frac{\phi^{-2}}{3}}},
\end{equation}

\begin{figure}[h!]
 \begin{center}$
 \begin{array}{cccc}
\includegraphics[width=50 mm]{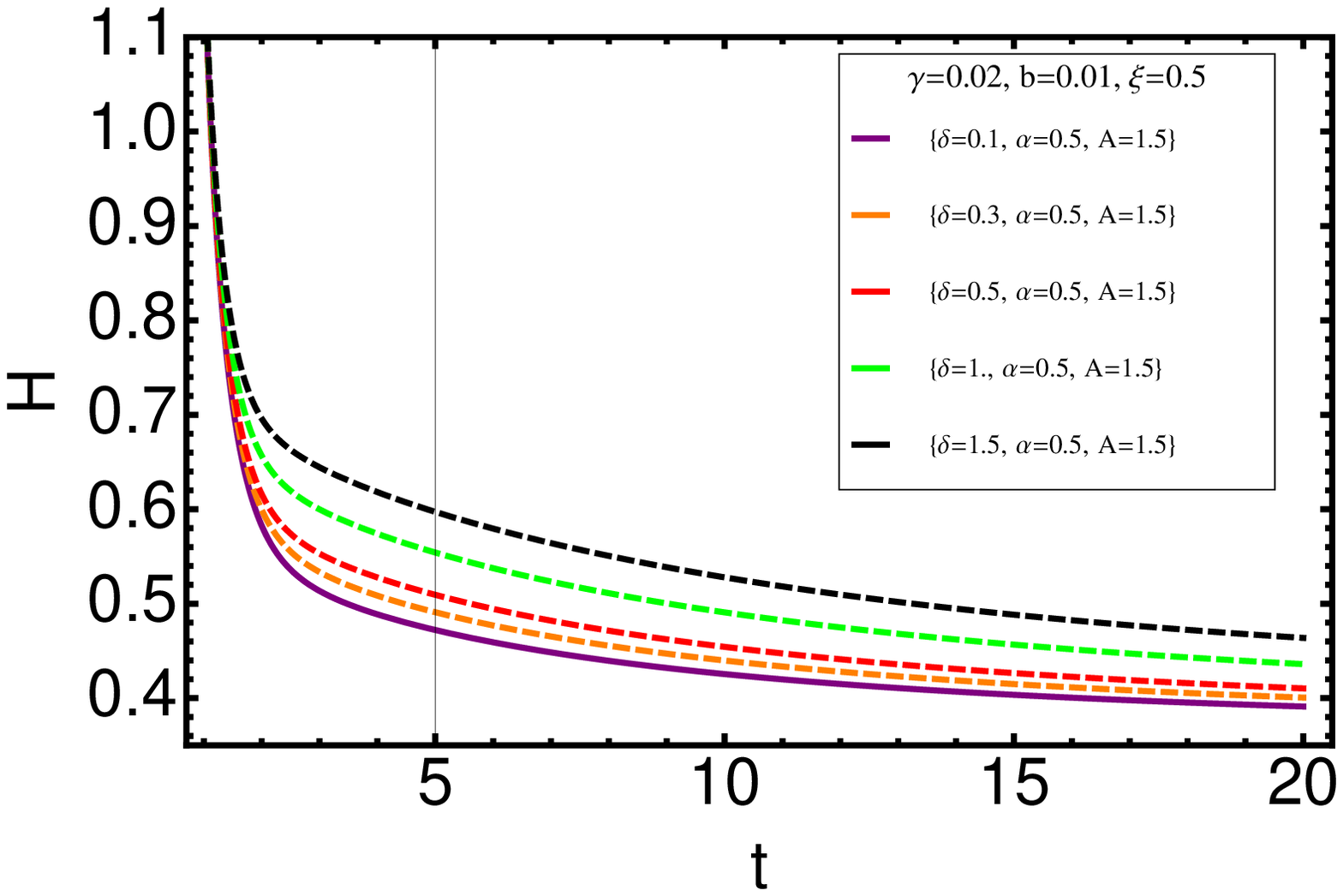} &
\includegraphics[width=50 mm]{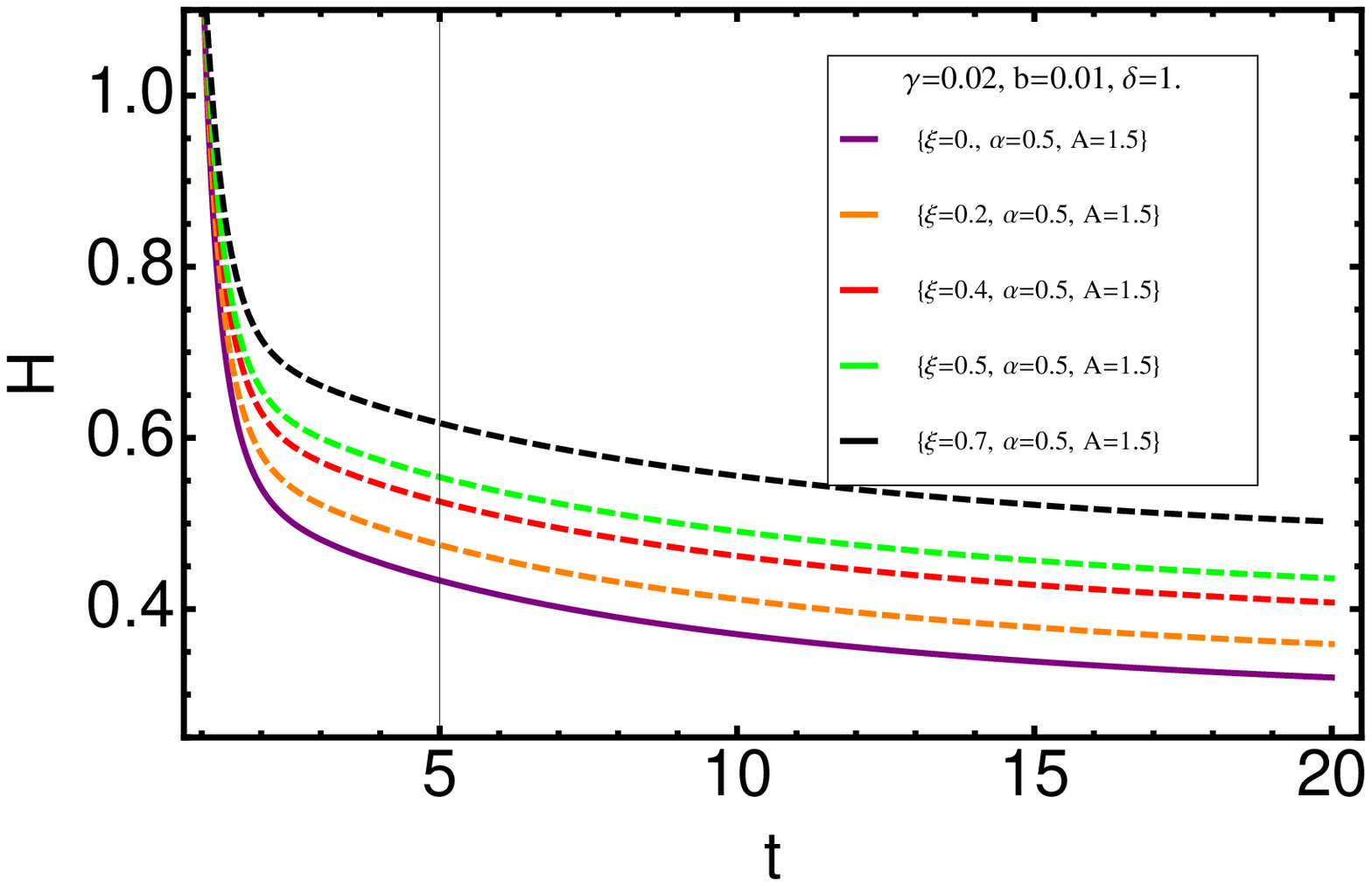}\\
\includegraphics[width=50 mm]{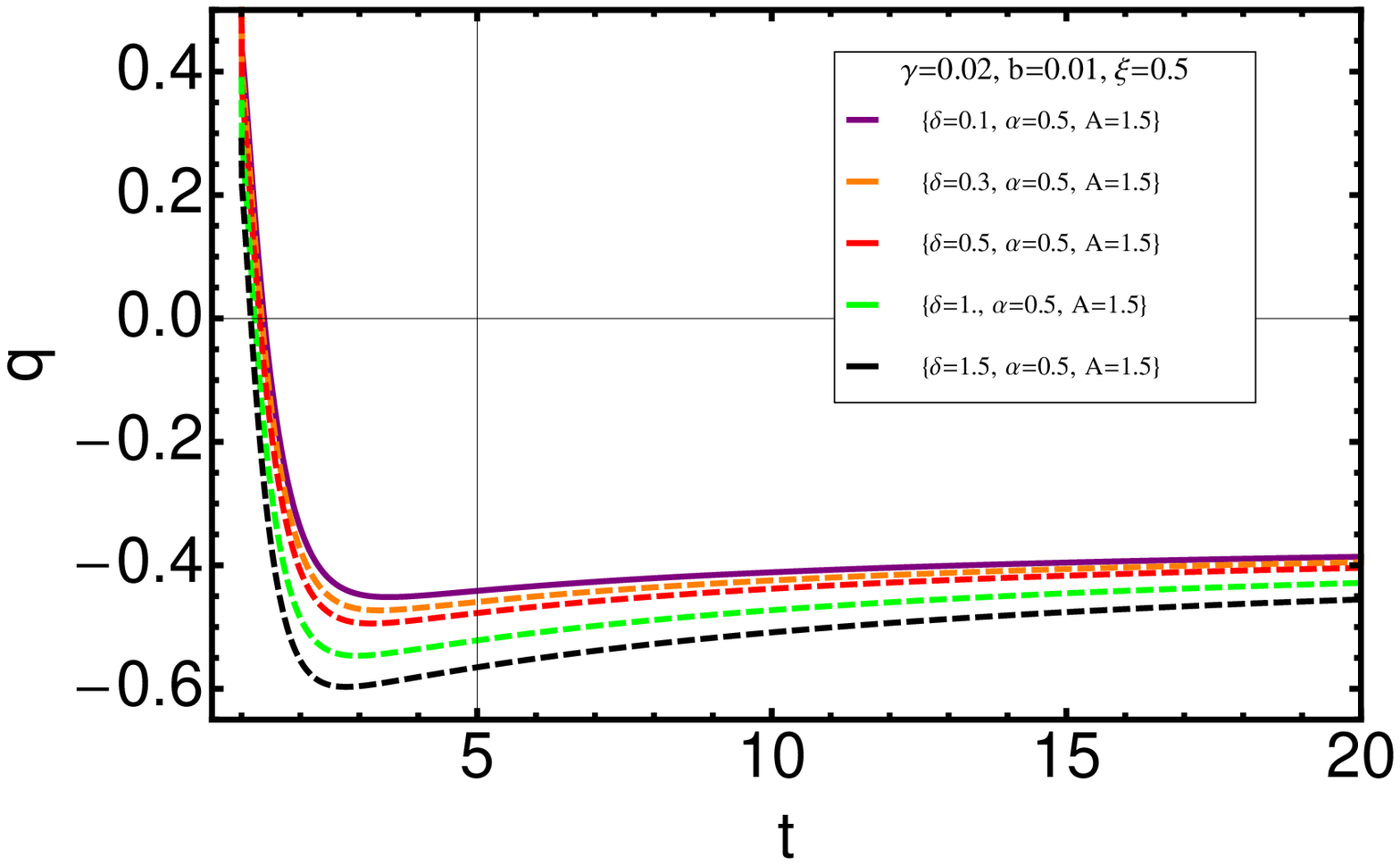} &
\includegraphics[width=50 mm]{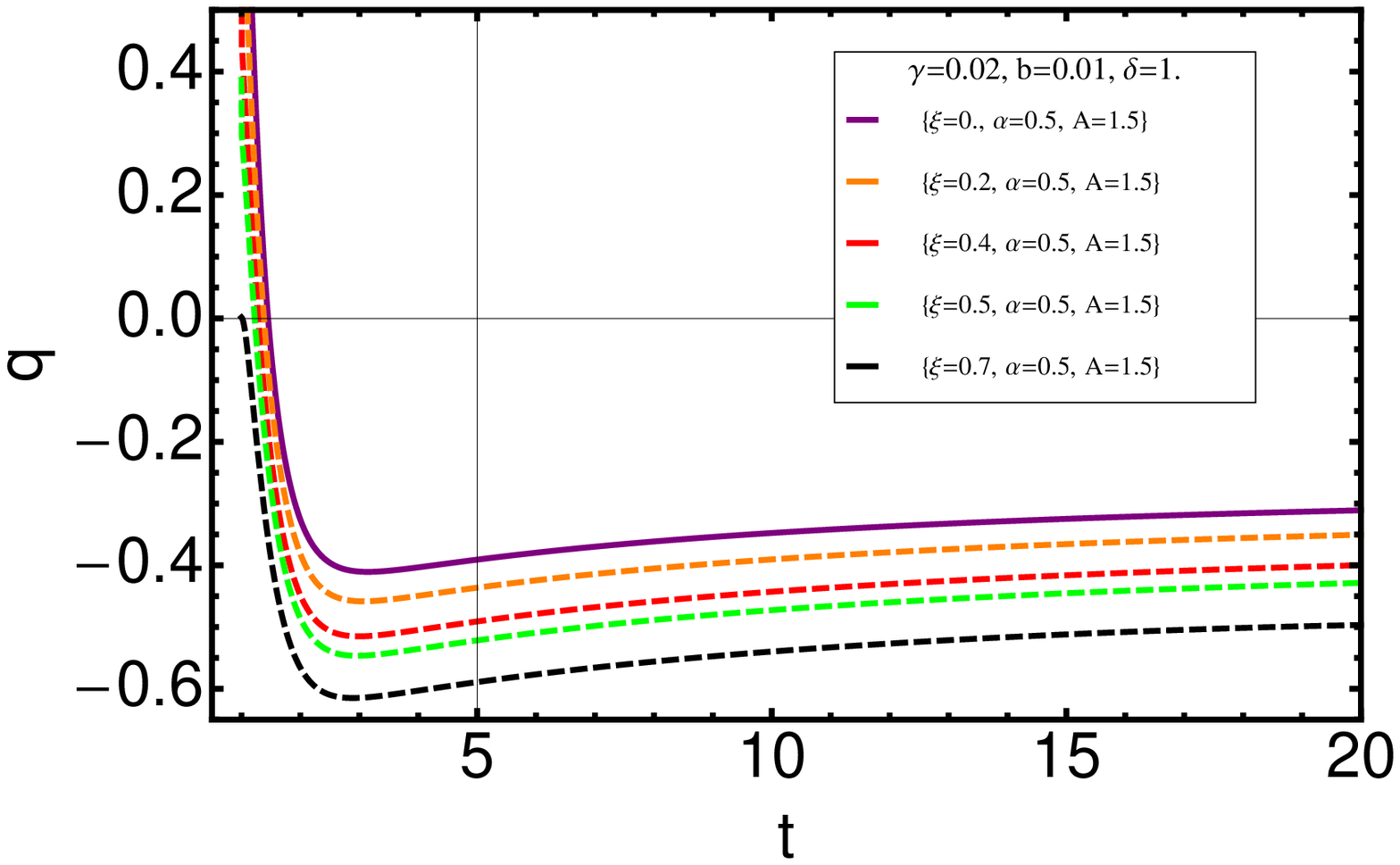}
\end{array}$
 \end{center}
\caption{Behavior of Hubble parameter $H$ and deceleration parameter $q$ against $t$ for varying $\Lambda$. Model 4.}
 \label{fig:1}
\end{figure}

\begin{figure}[h!]
 \begin{center}$
 \begin{array}{cccc}
\includegraphics[width=50 mm]{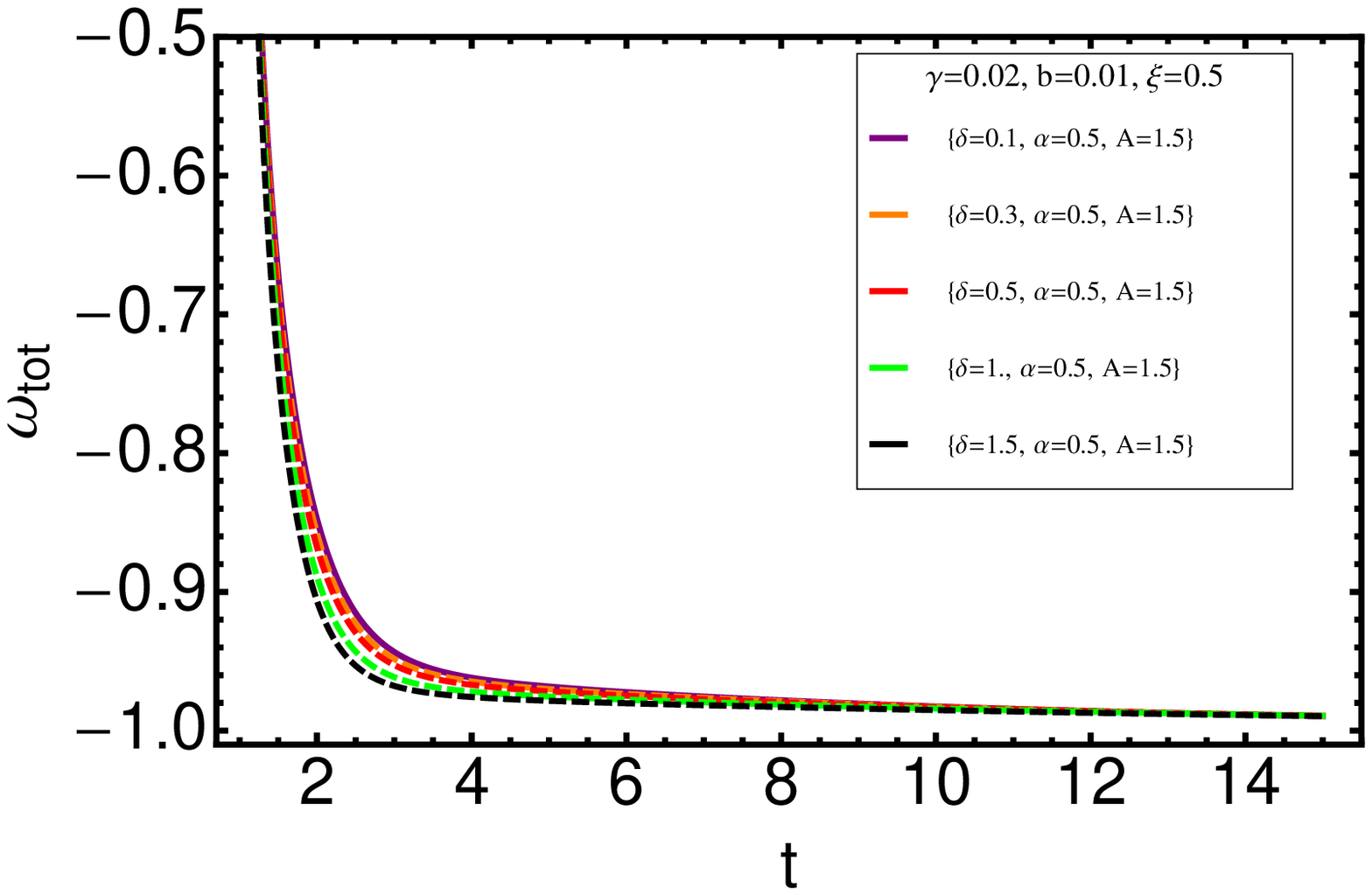} &
\includegraphics[width=50 mm]{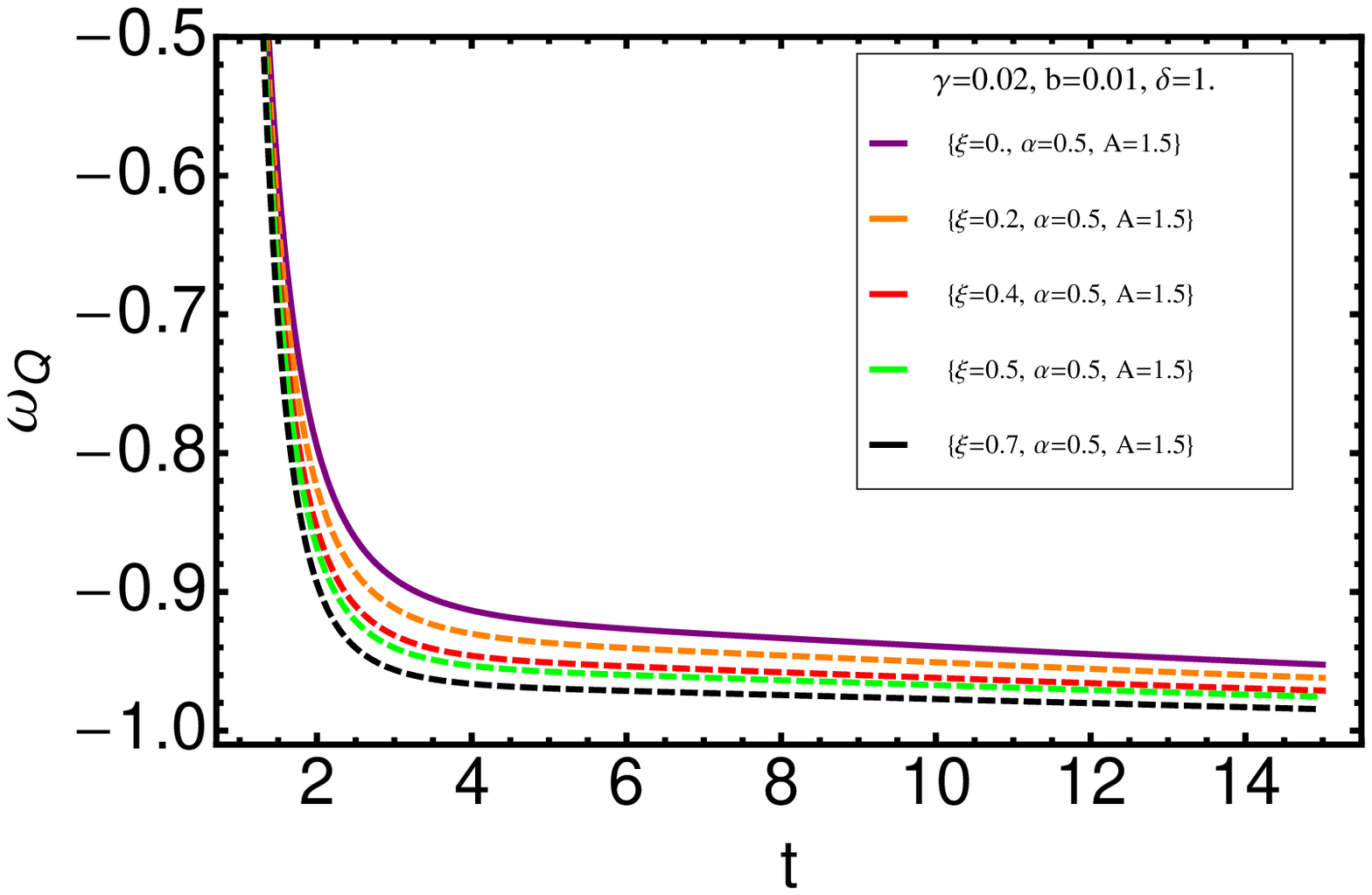}
 \end{array}$
 \end{center}
\caption{Behavior of the EoS parameters $\omega_{tot}$ and $\omega_{Q}$ against $t$ for the case corresponding to varying $\Lambda$. Model 4.}
 \label{fig:2}
\end{figure}

In Fig. 7, we can see that the Hubble expansion parameter is a decreasing function of time $t$, which yields to a constant at the late time as expected. It is clear that $\delta$ and $\xi$ increase the value of the Hubble expansion parameter but decrease value of the deceleration parameter. In order to obtain the deceleration parameter in agreement with observational data we should choose larger values of $\delta$ and $\xi$. Also acceleration to deceleration phase transition happen in this model. We find an instability at the initial time but the model is completely stable at the late time.\\
Fig. 8 shows that the EoS parameters yield to -1 at the late time in agreement with observational data. Also effects of $\delta$ and $\xi$ illustrated in the plots of Fig. 8.

\begin{figure}[h!]
 \begin{center}$
 \begin{array}{cccc}
\includegraphics[width=50 mm]{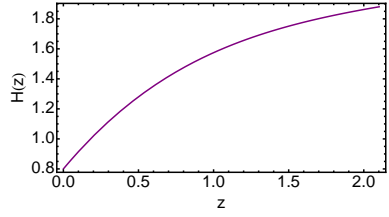}
 \end{array}$
 \end{center}
\caption{Hubble parameter $H$ against redshift $z$ for varying $\Lambda$. Model 4.}
 \label{fig:2-2}
\end{figure}

In Fig. 9, we can see behavior of the Hubble expansion parameter with the redshift which is also agree with observational data, since it is increasing function.

\subsection{\large{Model 5}}
A polytropic fluid interacting with the quintessence dark energy yield to the following differential equation,
\begin{equation}\label{eq:M2-2}
\dot{\rho}_{P}+3H \left (1+\omega_{P}-\frac{\gamma}{3H}\frac{\dot{\phi}}{\phi} \right )\rho_{P}=3H \left ( b-\frac{\gamma}{3H}\frac{\dot{\phi}}{\phi} \right )\rho_{Q}+9\xi H,
\end{equation}
with the Hubble parameter obtained as,
\begin{equation}\label{eq:Hubble-2}
H=\frac{1}{\sqrt{3}}\sqrt{\frac{\rho+\delta e^{[-\phi_{0}\phi]}+\beta^{2}}{1-\frac{\phi^{-2}}{3}}}.
\end{equation}
In Fig. 10, we can see behavior of the Hubble expansion parameter and the deceleration parameter with time and find that the value of $K$ decreases value of the Hubble expansion parameter while increases value of the deceleration parameter. Also, acceleration to deceleration phase transition seen in this model.\\
However, squared sound speed which plotted in Fig. 11 shows that this model similar to the constant $\Lambda$ is completely instable at the late time and will be useful only for the early universe.

\begin{figure}[h!]
 \begin{center}$
 \begin{array}{cccc}
\includegraphics[width=50 mm]{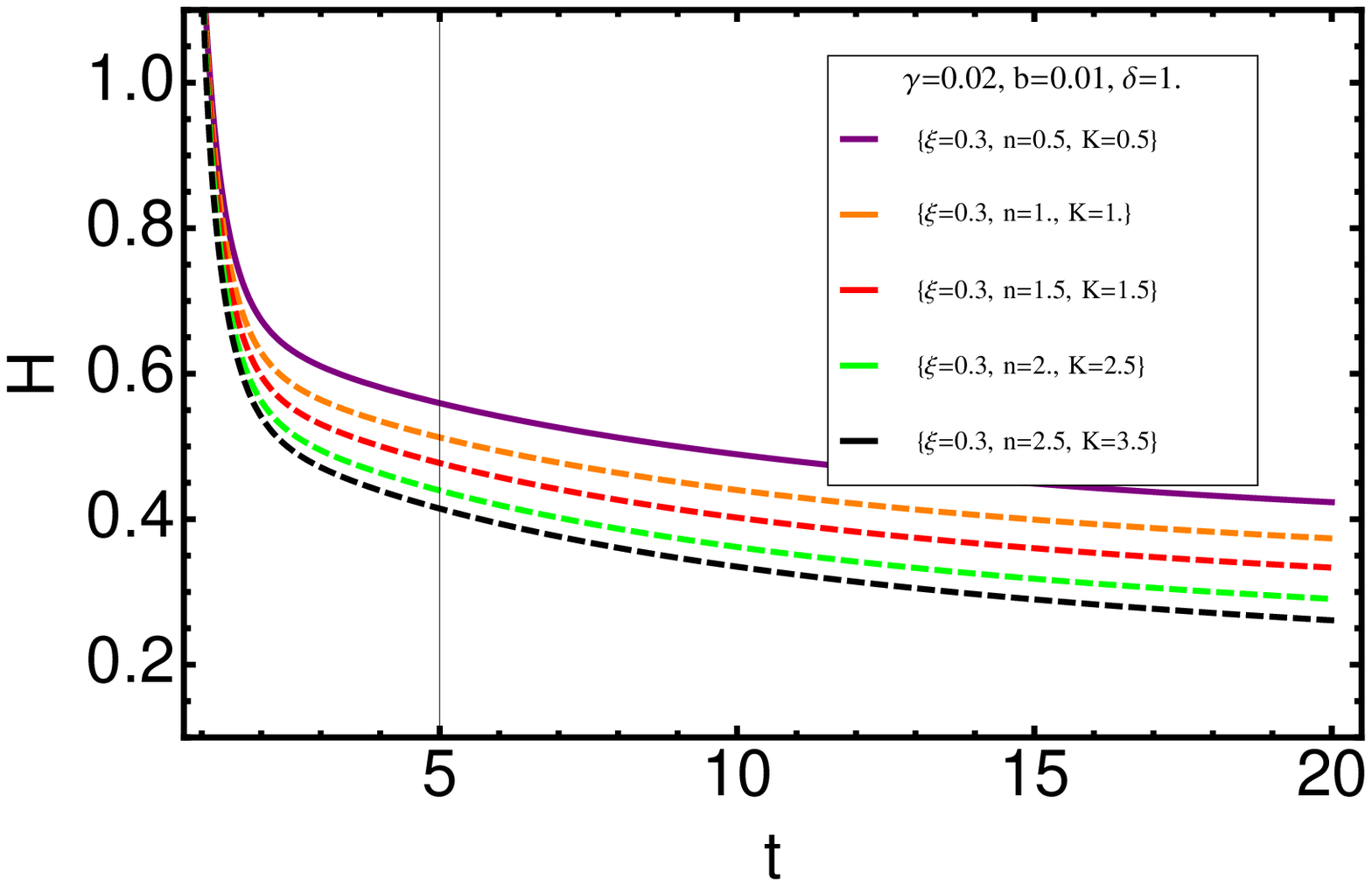} &
\includegraphics[width=50 mm]{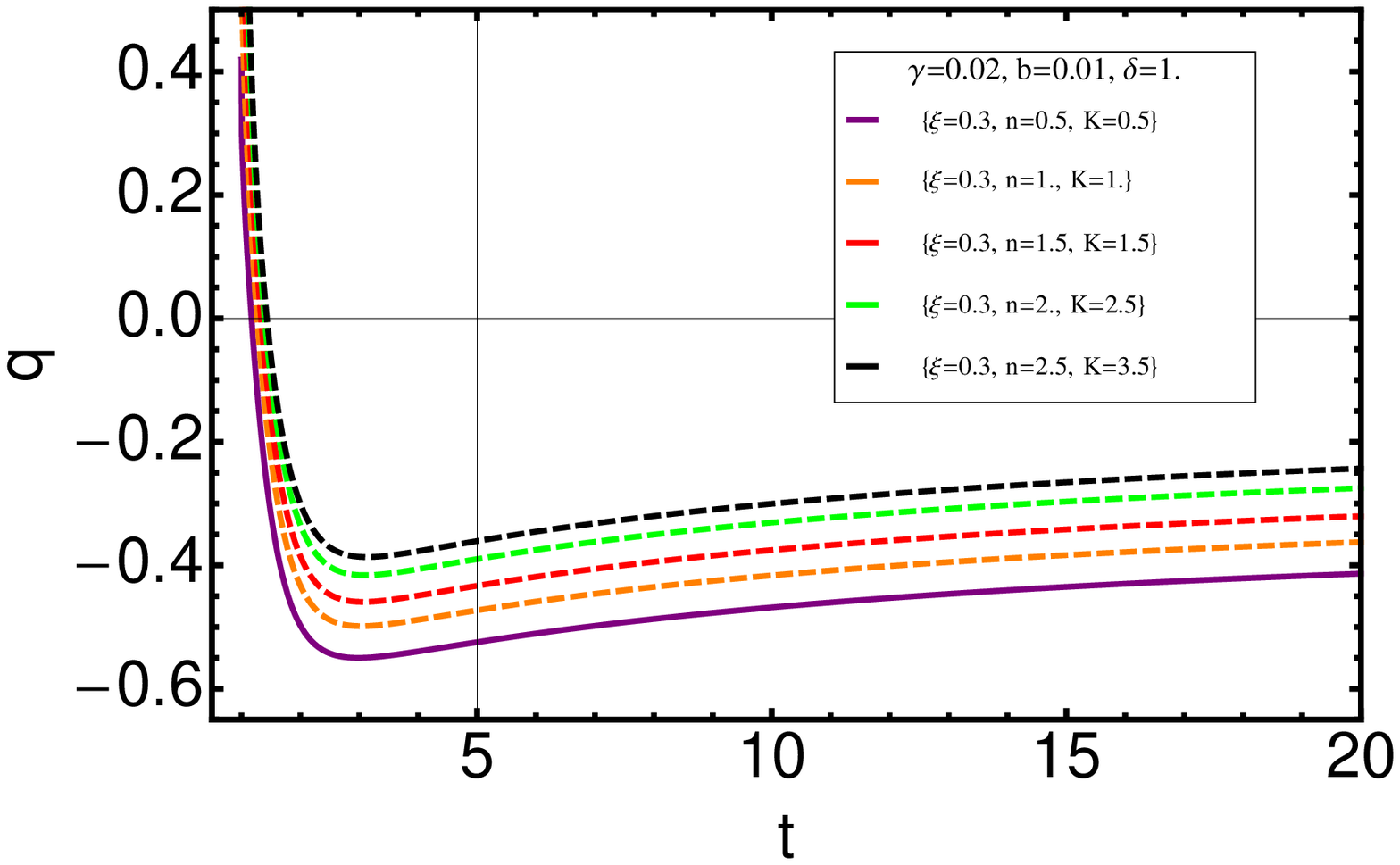}
 \end{array}$
 \end{center}
\caption{Behavior of Hubble parameter $H$ and deceleration parameter $q$ against $t$ for varying $\Lambda$. Model 5.}
 \label{fig:3}
\end{figure}

\begin{figure}[h!]
 \begin{center}$
 \begin{array}{cccc}
\includegraphics[width=50 mm]{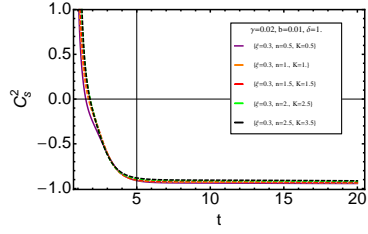}
 \end{array}$
 \end{center}
\caption{Squared sound speed against $t$  for varying $\Lambda$. Model 5.}
 \label{fig:4}
\end{figure}

\subsection{\large{Model 6}}
Finally, in the third model for the case of varying $\Lambda$ where a viscous van der Waals gas with EoS of the general form,
\begin{equation}\label{eq:EoS_M6}
P=\frac{A\rho_{W}}{B-\rho_{W}}-B\rho_{W}^{2}-3\xi H,
\end{equation}
interacts with the quintessence dark energy, the differential equations describing the dynamics of the energy densities of both components are given,
\begin{equation}\label{eq:rhoW-2}
\dot{\rho}_{W}+3H\rho_{W} \left ( 1+\frac{A}{B-\rho_{W}} -B\rho_{W} -\frac{\gamma}{3H}\frac{\dot{\phi}}{\phi} \right )=3H\rho_{Q}\left ( b-\frac{\gamma}{3H}\frac{\dot{\phi}}{\phi} \right )+9\xi H,
\end{equation}
and,
\begin{equation}\label{eq:rhoQ3}
\dot{\rho}_{Q}+3H\rho_{Q} \left ( 1+b+\omega_{Q} -\frac{\gamma}{3H} \frac{\dot{\phi}}{\phi} \right )=-\gamma \rho_{i}\frac{\dot{\phi}}{\phi},
\end{equation}
where $\omega_{Q}$ is the EoS parameter of the dark energy. We obtain the behavior of important cosmological parameters as illustrated in Fig. 12. Moreover, Fig. 13 shows that this model is also instable at the late time.\\

\begin{figure}[h!]
 \begin{center}$
 \begin{array}{cccc}
\includegraphics[width=50 mm]{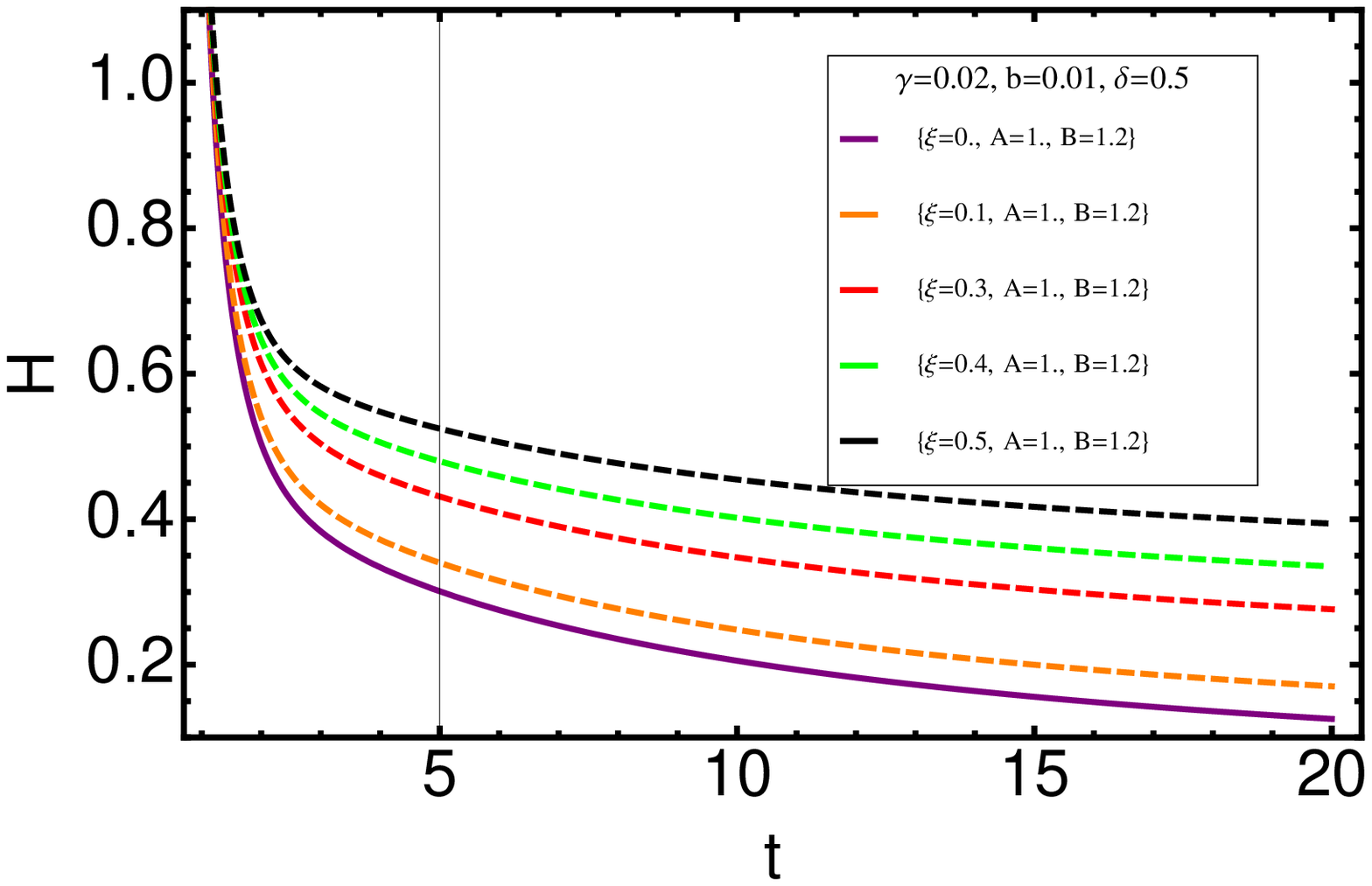} &
\includegraphics[width=50 mm]{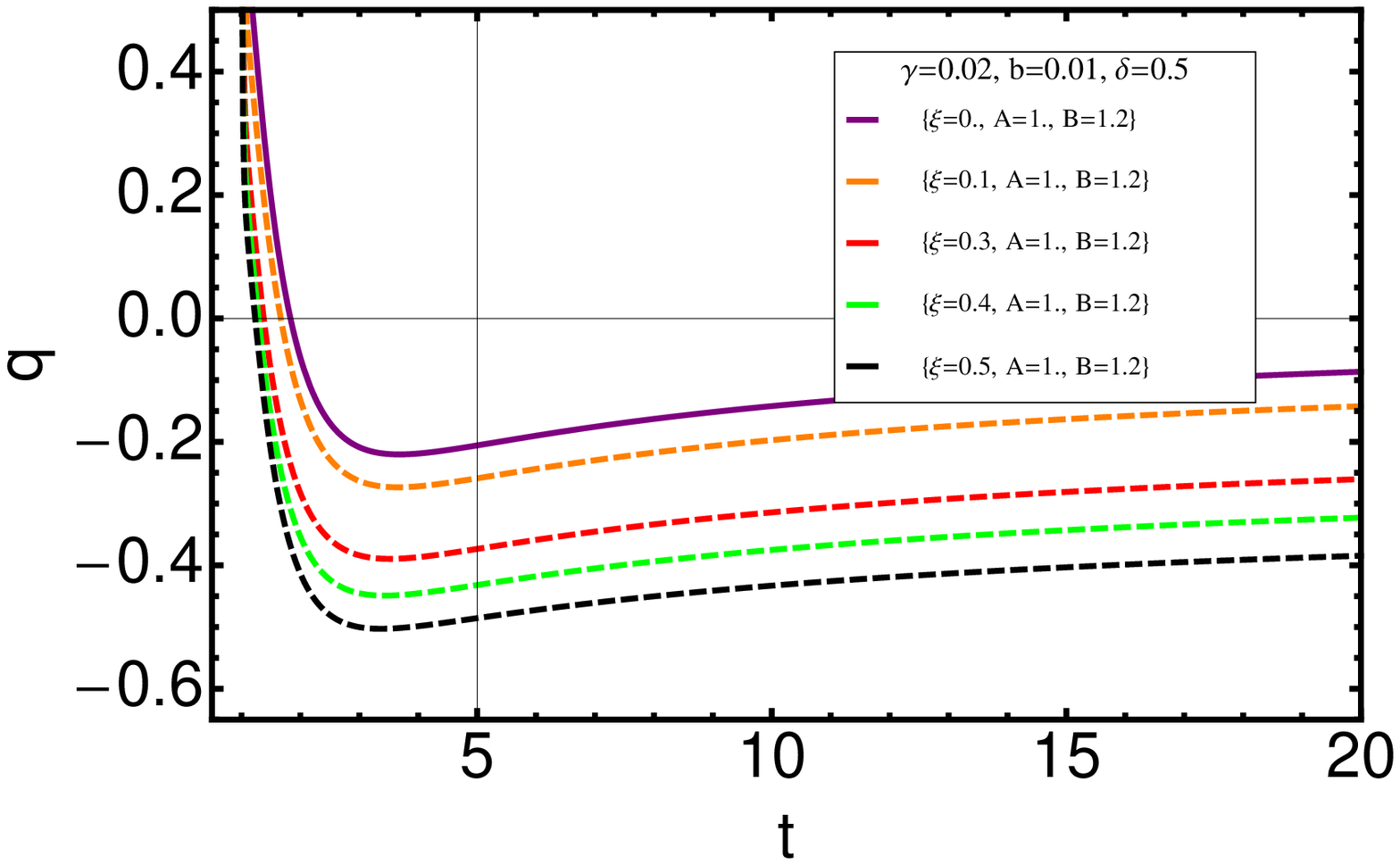}
 \end{array}$
 \end{center}
\caption{Behavior of Hubble parameter $H$ and deceleration parameter $q$ against $t$ for varying $\Lambda$. Model 6.}
 \label{fig:3-2}
\end{figure}

\begin{figure}[h!]
 \begin{center}$
 \begin{array}{cccc}
\includegraphics[width=50 mm]{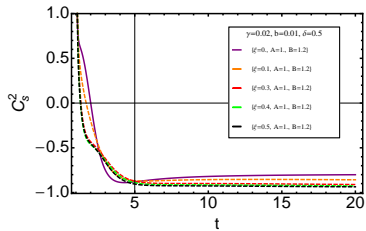}
 \end{array}$
 \end{center}
\caption{Squared sound speed against $t$  for varying $\Lambda$. Model 6.}
 \label{fig:4-2}
\end{figure}
Therefore we can choose the first model (model 4) as an appropriate model for the late time cosmology. Apart the instabilities which discussed above for the second and third models we can construct models by using more observational data which discussed in the next section.
\section{Observational constraints on interacting models with varying $\Lambda$}
The $SNIa$ test is based on the distance modulus $\mu$ which is related to the luminosity
distance $D_L$ by,
\begin{equation}
\mu=m-M=5\log_{10}{D_L},
\end{equation}
where $D_{L}$ defined as,
\begin{equation}
D_{L}=(1+z)\frac{c}{H_{0}}\int_{0}^{z}{\frac{dz'}{\sqrt{H(z')}}}.
\end{equation}
The quantities $m$ and $M$ denote the apparent and the absolute magnitudes, respectively. Baryonic acoustic oscillations have their origin in oscillations in the photon-baryon plasma at the moment of the decoupling at about $z = 1.090$. They can be characterized by the distance scale,
\begin{equation}
A=\frac{\sqrt{\Omega_{m0} } }{H(z_{b})^{1/3}} \left[ \frac{1}{z_{b}} \int_{0}^{z_{b}}{\frac{dz}{H(z)}} \right ]^{2/3}.
\end{equation}
The WiggleZ-data \cite{74} indicates the following information about $A$ and $z_{b}$: $A = 0.474 \pm 0.034$, $0.442 \pm 0.020$ and $0.424 \pm 0.021$ for the redshifts $z_{b} = 0.44$, $0.60$ and $0.73$, respectively. The key quantity of a statistical analysis is the $\chi^{2}$ parameter,
\begin{equation}
\chi^{2}{(x^{j})}=\sum_{i}^{n}\frac{(f(x^{j})_{i}^{t}-f(x^{j})_{i}^{0})^{2}}{\sigma_{i}},
\end{equation}
where $f(x^{j})_{i}^{t}$ is the theoretical evaluation of a given observable, depending on $x^{j}$ free parameters, $f(x^{j})_{i}^{0}$ is the corresponding observational value, and $n$ is the total number of observational data for the given test. There are many different $SNIa$ data sets, obtained with different techniques. In some cases, these different samples may give very different results. The second point is the existence of two different calibration methods. One of them uses cosmological relations and
takes into account $SNIa$ with high $z$, the other one, using astrophysical methods, is suitable for small $z$ (MLCS2k2). Our observational analysis of the background dynamics uses the following three tests: the differential age of old objects based on the $H(z)$ dependence as well as the data from $SNIa$ and from $BAO$. A fourth test could potentially be added: the position of the first peak of the anisotropy spectrum of the cosmic microwave background radiation (CMB). However, the CMB test implies integration of the background equations until $z \approx 1.000$ which requires the introduction of the radiative component. But the inclusion of such radiative component considerably changes the structure of the equations and no analytic expression for $H(z)$ is available. Hence, we shall limit ourselves to the mentioned three tests for which a reliable estimation is possible.\\
In the following tables we fix parameters of three models by using mentioned observational data.\\

\begin{center}
    \begin{tabular}{ | l | l | l | l | l | l | l | }
    \hline
    M & $\delta$ & $\gamma$ & $b$ & $\xi$  &  $H_{0}$ & $\Omega_{m0}$ \\ \hline
    4 & $1.3^{+0.2}_{-0.2}$ & $0.02^{+0.03}_{-0.01}$ & $0.01^{+0.02}_{-0.01}$ & $0.5^{+0.25}_{-0.45}$ & $1.1^{+0.1}_{-0.2}$ &  $0.3^{+0.15}_{-0.15}$ \\ \hline
    5 & $1.4^{+0.2}_{-0.3}$ & $0.02^{+0.02}_{-0.01}$ & $0.01^{+0.02}_{-0.01}$ & $0.3^{+0.35}_{-0.15}$ & $0.8^{+0.2}_{-0.3}$ &  $0.25^{+0.3}_{-0.1}$ \\ \hline
    6 & $0.75^{+0.35}_{-0.15}$ & $0.02^{+0.03}_{-0.02}$ & $0.03^{+0.01}_{-0.02}$ & $0.35^{+0.15}_{-0.1}$ & $1.4^{+0.05}_{-0.1}$ &  $0.23^{+0.03}_{-0.03}$ \\ \hline
    \end{tabular}
\end{center}

\begin{center}
    \begin{tabular}{ | l | l | l | l | l | l |}
    \hline
    M & $A$ & $B$ & $\alpha$ & $K$  &  $n$ \\ \hline
    4 & $1.7^{+0.2}_{-0.3}$ & $0.3^{+0.05}_{-0.15}$ & $0.5^{+0.2}_{-0.4}$ & $-$ & $-$ \\ \hline
    5 & $-$ & $-$ & $-$ & $1.5^{+0.3}_{-0.5}$ & $1.2^{+0.5}_{-0.3}$ \\ \hline
    6 & $1.2^{+0.3}_{-0.1}$ & $1.1^{+0.4}_{-0.2}$ & $-$ & $-$ & $-$ \\ \hline

    \end{tabular}
\end{center}

In Fig. 14, we can see the behavior of $\mu$ in all models, which is approximately in agreement with observational data.

\begin{figure}[h!]
 \begin{center}$
 \begin{array}{cccc}
\includegraphics[width=50 mm]{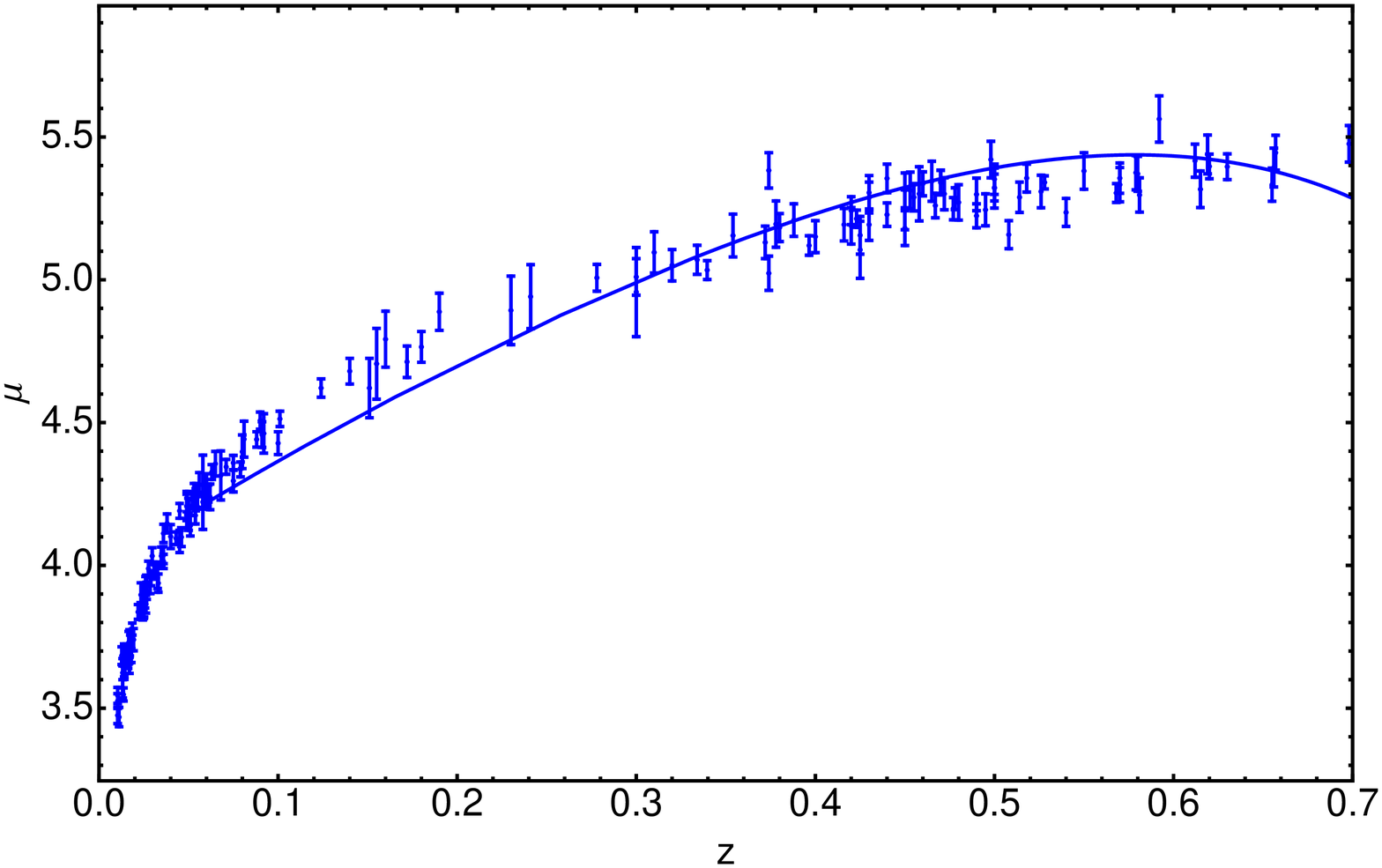}&
\includegraphics[width=50 mm]{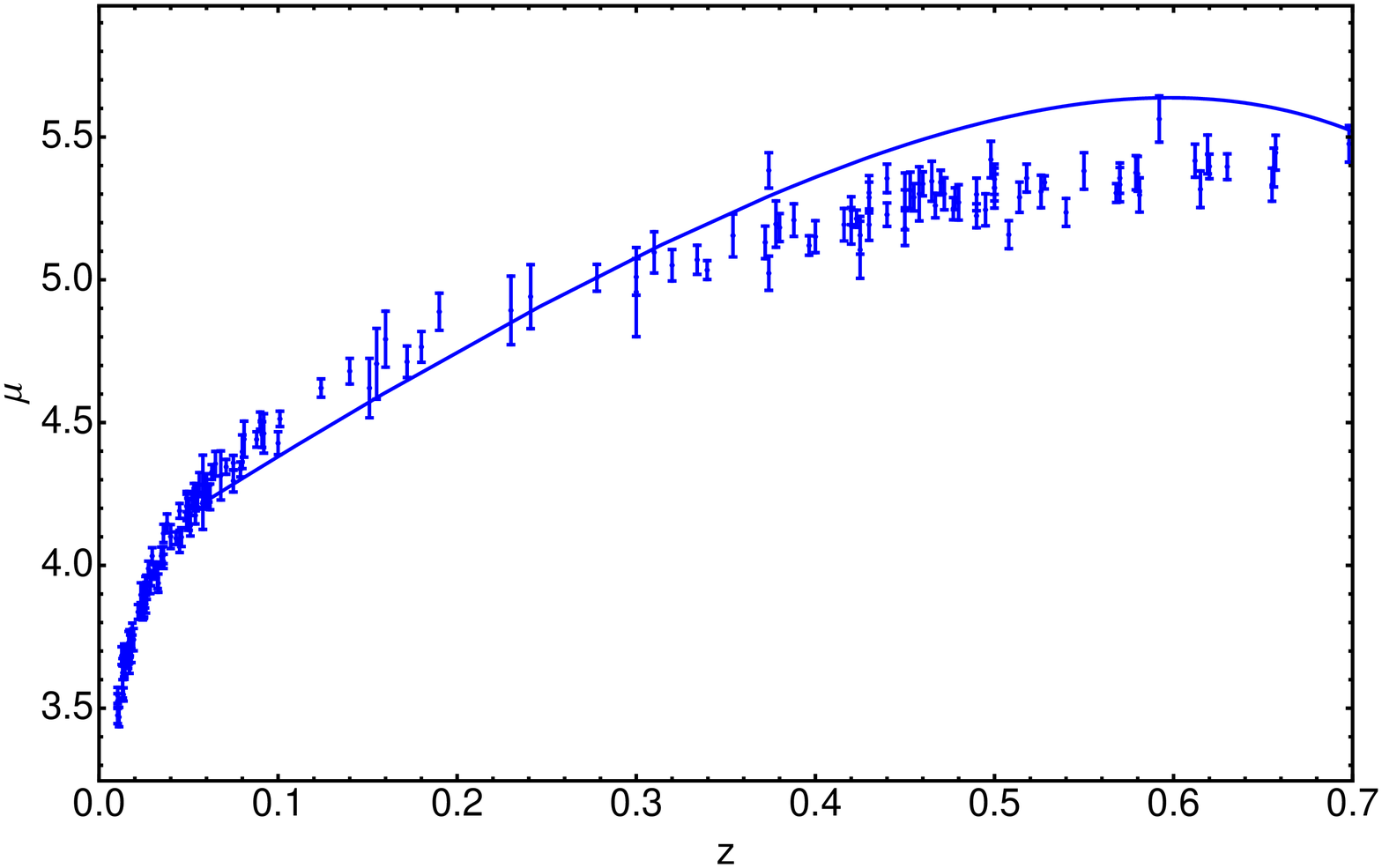} &
\includegraphics[width=50 mm]{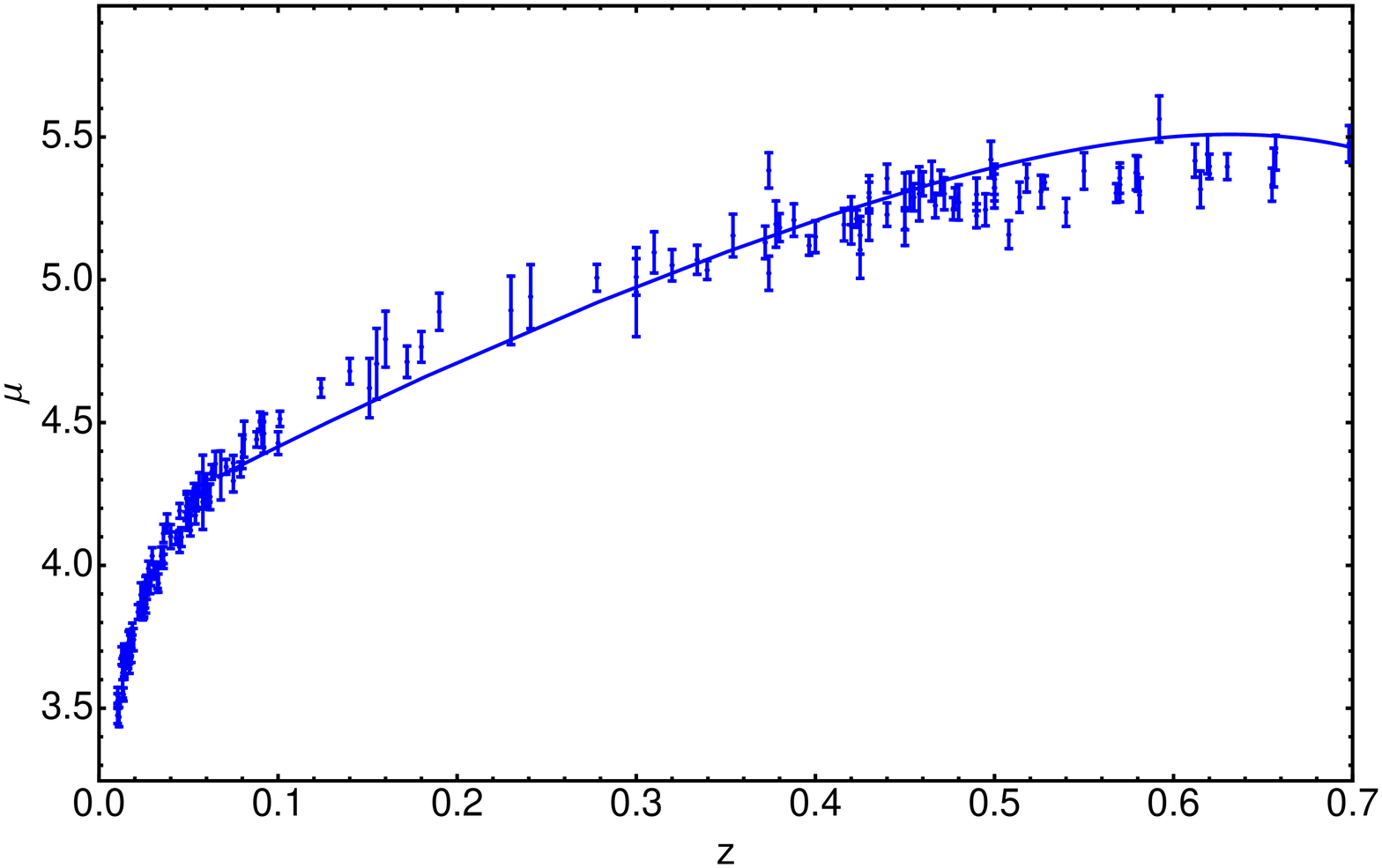}
 \end{array}$
 \end{center}
\caption{Observational data (SneIa+BAO+CMB) for distance modulus versus our theoretical results for varying $\Lambda$ in the models 4, 5 and 6.}
 \label{fig:muz}
\end{figure}

\section{Conclusion}
In this paper we considered three different cosmological models for the universe based on Lyra geometry. First of all we introduced our models and then obtained field equations which solved numerically to find behavior of cosmological parameters. In order to find effect of varying $\Lambda$ we also studied the case of constant $\Lambda$ and found that presence of $\Lambda$ is necessary to obtain results in agreement with observational data. We assumed viscous modified Chaplygin gas (model 1 and 4), viscous polytropic gas (model 2 and 5) or viscous Van der Waals gas (model 3 and 6) as a component which at the early universe play the role of dark matter with $\omega=0$, but at late times it tends to a cosmological constant. Moreover, we have a quintessence field, which will contribute to the dark energy sector including possibility of interaction between components. Easily we can check that $\Omega_{DE}$ and $\Omega_{DM}$ are of the same order. Also we considered case of varying $\Lambda$ and studied behavior of cosmological parameters numerically. We used observational data to fix parameters of the models and seen agreement with observational data by investigation of $H(z)$. By using stability analysis we concluded that the model 2 is the best model considered in this paper to describe universe.\\
For the future work we can extend presence paper to include shear viscosity or varying bulk viscosity \cite{68}, also we can consider cosmic Chaplygin gas versions \cite{69} to obtain more general model.

\end{document}